\documentclass[aoas,preprint]{imsart}

\RequirePackage[OT1]{fontenc}
\RequirePackage[utf8]{inputenc}
\RequirePackage{amsthm,amsmath}
 
\usepackage{graphicx}
\usepackage{amsthm,amsmath,amssymb}

\usepackage{natbib}

\usepackage[ruled]{algorithm2e}
\usepackage{url}

\usepackage[usenames,dvipsnames]{color}
\usepackage[colorlinks=true]{hyperref}
\hypersetup{
  colorlinks,
 citecolor=blue,
 linkcolor=red,
 urlcolor=Violet}

\newcommand{\AB}{\allowbreak}
\newcommand{\bx}{\bold{x}}
\newcommand{\bX}{\bold{X}}
\newcommand{\bz}{\bold{z}}
\newcommand{\bZ}{\bold{Z}}

\newcommand{\bmu}{\boldsymbol{\mu}}
\newcommand{\bpi}{\boldsymbol{\pi}}
\newcommand{\balpha}{\boldsymbol{\alpha}}

\newcommand{\bdelta}{\boldsymbol{\delta}}
\newcommand{\bgamma}{\boldsymbol{\gamma}}

\numberwithin{equation}{section}
\numberwithin{figure}{section}

\allowdisplaybreaks[1]

\startlocaldefs

\endlocaldefs

\begin{document}

\begin{frontmatter}

\title{Opinion mining from twitter data using evolutionary multinomial mixture models}
\runtitle{Opinion mining from twitter data}

\begin{aug}

\affiliation{Laboratoire ERIC, Universit\'e de Lyon - Lumière \thanksmark{m1}}

\author{\fnms{Md. Abul} \snm{Hasnat}\thanksref{m1}\ead[label=e1]{mhasnat@gmail.com}},
\author{\fnms{Julien} \snm{Velcin}\thanksref{m1}\ead[label=e3]{julien.velcin@univ-lyon2.fr}},
\author{\fnms{Stephane} \snm{Bonnevay}\thanksref{m1}\ead[label=e2]{stephane.bonnevay@univ-lyon1.fr}}
\and \author{\fnms{Julien} \snm{Jacques}\thanksref{m1}\ead[label=e4]{julien.jacques@univ-lyon2.fr}}

\runauthor{Hasnat et al.}

\end{aug}

\begin{abstract}
\emph{Image} of an entity can be defined as a structured and dynamic representation which can be extracted from the opinions of a group of users or population. Automatic extraction of such an image has certain importance in political science and sociology related studies, e.g., when an extended inquiry from large-scale data is required.
We study the images of two politically significant entities of France. These images are constructed by analyzing the opinions collected from a well known social media called \textit{Twitter}. Our goal is to build a system which can be used to automatically extract the image of entities over time. 

In this paper, we propose a novel evolutionary clustering method based on the parametric link among Multinomial mixture models. 
First we propose the formulation of a generalized model that establishes parametric links among the Multinomial distributions. Afterward, we follow a model-based clustering approach to explore different parametric sub-models and select the best model. For the experiments, first we use synthetic temporal data. Next, we apply the method to analyze the annotated social media data. Results show that the proposed method is better than the state-of-the-art based on the common evaluation metrics. Additionally, our method can provide interpretation about the temporal evolution of the clusters. 
\end{abstract}
\end{frontmatter}

\section{Introduction}
\label{sec:intro}
We define an \textit{image} as a multi-faceted representation that aggregates a set of opinions or general impressions regarding an entity. By entity, we mean a politician, a celebrity, a company, a brand, etc.
In this research, we are particularly interested to use annotated social media data to extract the image of two French politicians and observe its changes/evolution over time. We consider the annotated data from the \textit{ImagiWeb} project  \citep{velcin2014investigating} which are extracted before and after the 2012 French presidential election. The annotation provides a compact and meaningful representation for each tweet.
Our goal is to develop a temporal/evolutionary clustering technique, which groups the annotated opinions and then extracts the image of an entity over time from the clustering results. Subsequently, we want to explain/interpret the temporal changes of the image created from each group of users.

In the recent years, the social media plays a significant role in many aspects of our daily activity.
There exist numerous popular social media such as Twitter or Facebook, where the users (people) often provide their opinions about particular entity, e.g., persons (politician, actor), products consumed in the daily life, etc. A common method to analyze such data is to use a clustering method that naturally groups the users/opinions, and then investigate each group independently. An important property of these data is that they may change \textit{over time} due to changes of the attributes, and appearance/disappearance of users. Moreover, users may change their opinion about the targeted entity. 

An ordinary clustering method is unlikely to adapt with such temporal dynamics of the data, as it does not consider any relevant information such as history and temporal effects. The notion of evolutionary clustering \citep{chakrabarti2006evolutionary, xu2014adaptive, chi2009evolutionary, xu2012generative} appears in such situations, where the method should be specialized in clustering temporal data by taking care of the historic information and current data altogether. Numerous methods exist, which address these issues appropriately and cluster temporal data. These methods are based on different strategies, such as spectral clustering \citep{chi2009evolutionary, xu2014adaptive} and probabilistic generative model \citep{blei2006dynamic, xu2012generative, kim2015temporal}. However, it remains an important issue - how to interpret the evolution of the clusters. In this research, we are motivated by this issue and propose a novel method based on the Multinomial mixture model \citep{bishop2006pattern} to cluster the temporal data as well as interpret the evolution of the clusters through some prior belief. Therefore, we propose a novel method which simultaneously performs evolutionary clustering and interpreting the evolution.

Multinomial Mixture (MM) model based clustering strategy is a popular method for clustering discrete data \citep{meilua2001experimental, silvestre2014identifying, hasnat2015mbhcfmm, agresti2002categorical}. Most recently, it has been exploited to perform evolutionary clustering \citep{kim2015temporal}. In this research, we consider MM as the core model for the data and propose an evolutionary clustering method by deriving appropriate link between the parameters of MM at different time.

Parametric link among probability distributions has been used in the context of transfer learning \citep{biernacki2002generalized, jacques2010extension, beninel2012parametric}, where the goal is to adapt a clustering model from a source population to a target one. In the context of continuous features, \cite{biernacki2002generalized} proposed a parametric link between the Normal distributions. \cite{jacques2010extension} extended it for the binary features using Bernoulli distribution. However, no such formulation exists for the Multinomial distribution. Moreover, such parametric link-based methods are never considered in the context of evolutionary clustering. This research addresses both of these issues. 

This research proposes a novel evolutionary clustering method for extracting \textit{image} of political entities. The highlights of our contributions include: (a) propose a formulation for a parametric link among Multinomial distributions; (b) develop a novel evolutionary clustering method by exploiting the link parameters and (c) provide interpretation of the link parameters to interpret cluster evolutions.  
First, we use synthetic data to evaluate and compare the proposed method w.r.t. the state-of-the-art methods. Next, we apply it to analyze the temporal dynamics of social media data obtained from the \textit{ImagiWeb} project \citep{velcin2014investigating}. Results in Sec. \ref{sec:exp_res} show that the proposed method is better than the state-of-the-art methods.

In the rest of the paper, we present the data in Sec. \ref{sec:the_data_iw}, describe our proposed method in Sec. \ref{sec:pl_ecl}, present the experimental results in Sec. \ref{sec:exp_res}, provide analysis of the political data in Sec. \ref{sec:anal_vis_int_real_data},
and finally draw conclusions in Sec. \ref{sec:discussion}.
\section{The Imagiweb project and the political opinion dataset}
\label{sec:the_data_iw}
%
%
We collected data from the $\textit{political opinion dataset}$ of the ImagiWeb\footnote{http://mediamining.univ-lyon2.fr/velcin/imagiweb/dataset.html} (IW-POD) project, see \cite{velcin2014investigating} for further details of data collection,  relevant statistics and representation.
IW-POD consists of manually annotated tweets, from May 2012 to January 2013, related to two French politicians: Francois Hollande (FH) and Nicolas Sarkozy (NS). First, these tweets are annotated into 11 different aspects, such as Attribute (Att), Person (Per), Entity (Ent), Skills (Skl), Political line (Pol), Balance (Bal), Injunction (Inj), Projet (Pro), Ethic (Eth), Communication (Com) and No aspect detected (N/A). Afterward, each aspect is annotated with 6 opinion polarities, such as very negative (-2), negative (-1), no polarity (0), Null, positive (+1) and very positive (+2). 
For example, the tweet - \textit{Sarko is more rational (orig: Sarko est plus rationnel)} is annotated with the aspect called \textit{Person} and polarity $+1$. It is about NS and indicates that the user provides positive opinion with an emphasis on the personal attribute. Another example, the tweet - \textit{Nicolas Sarkozy, the worst president of the Fifth Republic (Orig: Nicolas Sarkozy, le plus mauvais président de la Vème République)} is annotated with the aspect called \textit{Skill} and polarity $-1$. It is a negative opinion about NS and indicates that the user emphasizes on the skill of NS. 

In order to use these tweets for clustering, they are regrouped within the specified time epoch. Moreover, similar polarities are merged, e.g., two positives (+1 and +2) are merged into one as only positive (+). Therefore, each aspect consits of four polarities, such as positive (+), negative (-), zero (0) and undefined/null ($\emptyset$). As a consequence, finally each regrouped tweet represents the opinion of an user about a particular politician which is a $44 \, (11\times 4)$ dimensional vector of discrete data. 
In our experiment, we group opinions from IW-POD into three time\footnote{The first round of the presidential election was held in 22/04/2012 and the second round run-off was held on 06/05/2012. Therefore, the data collected during this election period belong to time epoch $t1$.} epochs: $t1$, $t2$ and $t3$, see Table \ref{tab:iw_pod_details} for details of the temporal data. Moreover, since the true number of clusters is unknown, we run clustering for different numbers of clusters ranging from 3 to 9. 

\begin{table}[h]
\centering
\caption{Details of the IW-POD dataset which is divided into three time periods. Each observation consists of a 44 dimensional discrete valued vector that encodes information about 11 different aspects each having 4 polarities.}
\begin{tabular}{|c|c|l|c|c|}
\hline
\textbf{\begin{tabular}[c]{@{}c@{}}Time \\ stamp\end{tabular}} & \textbf{\begin{tabular}[c]{@{}c@{}}Time \\ period\end{tabular}} & \textbf{Significance} & \textbf{\begin{tabular}[c]{@{}c@{}}Num. opinions \\ N. Sarkozy\end{tabular}} & \textbf{\begin{tabular}[c]{@{}c@{}}Num. opinions \\ F. Hollande\end{tabular}} \\ \hline
\textbf{t1}                                                    & 03/12 - 06/12                                                   & \begin{tabular}[x]{@{}c@{}}Before and \\ After Election\end{tabular}       & 1018                                                            & 1168                                                            \\ \hline
\textbf{t2}                                                    & 07/12 - 10/12                                                   & After Election        & 1067                                                            & 1079                                                            \\ \hline
\textbf{t3}                                                    & 11/12 - 01/13                                                   & After Election        & 1079                                                            & 708                                                             \\ \hline
\end{tabular}
\label{tab:iw_pod_details}
\end{table}
%
\section{Parametric Link Based Evolutionary Clustering}
\label{sec:pl_ecl}
We adopt the parametric link approach \citep{biernacki2002generalized, jacques2010extension} for evolutionary clustering by assuming that the source samples are equivalent to the samples at time epoch $t$ and target samples represent sample of time $t+1$. With this assumption, we incorporate linear link between Multinomials at different time epoch. The algorithm for the proposed clustering method is presented in Algorithm \ref{algo:pl_ecl}.
\subsection{Related work}
\label{sec:rel_work}
Evolutionary Clustering (ECL), also called \textit{clustering over time}, aims to cluster the data that dynamically evolves over time \citep{chakrabarti2006evolutionary}. Ordinary clustering methods are not appropriate as they group/partition the data samples only based on the certain properties of the data. In contrary, ECL methods cluster the data by additionally considering the temporal smoothness to reflect the long-term trends of the data while being robust to the short-term variations \citep{chakrabarti2006evolutionary, xu2014adaptive, chi2009evolutionary}. ECL should maintain four properties \citep{chakrabarti2006evolutionary} such as consistency, noise removal, smoothing and cluster correspondence. The demand and application of such clustering method are increasing rapidly due to the significant growth of the dynamic data in numerous domains. It has been successfully applied to analyze news \citep{xu2012generative}, social media \citep{kim2015temporal}, stock price \citep{xu2014adaptive}, photo-tag pairs \citep{chakrabarti2006evolutionary}, and documents \citep{blei2006dynamic}.

Temporal/evolutionary data clustering has been addressed from several viewpoints in the literature, which naturally raises several task-specific notions about ECL. A distinction among them can be as follows: (1) clustering (2) monitoring and (3) interpreting. In the following paragraphs, we review relevant literature based on this distinction.

Following the definition of \cite{chakrabarti2006evolutionary}, the ECL method clusters data by considering the historic information and current data. Based on this definition, in this research we do not consider the methods which do not take into account the historic information. Besides, in order to limit our focus on the parametric methods, we do not consider the methods from non-parametric Bayesian based approaches \citep{xu2008dirichlet, dubey2013nonparametric, kharratzadeh2015bayesian}. 

Numerous methods based on different techniques have been proposed in the literature \citep{chakrabarti2006evolutionary, xu2014adaptive, chi2009evolutionary, xu2012generative, kim2015temporal, blei2006dynamic}. \cite{chakrabarti2006evolutionary} provided a generic framework for this problem and proposed evolutionary version of k-means and hierarchical agglomerative clustering methods. Their proposed framework is based on optimizing a global cost function that consists of snapshot (static clustering) quality and history cost (temporal smoothness). This is considered as the first work for evolutionary clustering and has been subsequently extended by other researchers. \cite{chi2009evolutionary} proposed two evolutionary clustering methods based on spectral clustering strategy. In their approach, they added terms within the clustering cost functions in order to regularize the temporal smoothness.  \cite{xu2014adaptive} recently proposed AFFECT, which performs adaptive evolutionary clustering by estimating an optimal smoothing parameter. This approach is extended with several static clustering methods, such as k-means, hierarchical and spectral. A common property of these methods is that they specialized for continuous data and hence may not be an appropriate choice for clustering categorical data that is our concern in this research.

Dynamic Topic Model (DTM) is a well-known probabilistic method for analyzing temporal categorical data \citep{blei2006dynamic}. It was originally developed to analyze time evolution of topics in large document collections. DTM extends the popular topic modeling method called Latent Dirichlet Allocation (LDA) \citep{blei2003latent}. It uses Dirichlet prior based smoothing, which sometime over-smooth the data. As a consequence, it may cluster the data samples with non co-occurring features in the same group \citep{kim2015temporal}. This eventually causes DTM to underperform to cluster some classical non-textual temporal categorical data. Recently,  \cite{kim2015temporal} address this issue and proposed a probabilistic generative model based evolutionary clustering method, called Temporal Multinomial Mixture (TMM). TMM extends the classical Multinomial Mixture (MM) model by incorporating temporal dependency into the relation between data components of current time epoch and the clusters of the previous time epoch. MM is a well-known standard probabilistic model, which has been widely used to cluster static discrete/categorical data \citep{meilua2001experimental, silvestre2014identifying}. Similar to MM, TMM estimates model parameters using an Expectation Maximization (EM) algorithm. Although both DTM and TMM provide reasonable results to cluster temporal categorical data, they are unable to detect and provide any interpretation of the cluster evolutions, which is one of the main foci of this research. Indeed, TMM is more related to our proposed approach as we aim to establish parametric link among MMs at different time epochs.

The evolution monitoring task \citep{spiliopoulou2006monic, oliveira2010mec, ferlez2008monitoring, lamirel2012new} tracks the evolution of clusters by identifying the birth, death, split, merge and survival of clusters at different time. An external clustering method is first used at each time to cluster the data, e.g., \cite{spiliopoulou2006monic} and \cite{oliveira2010mec} used the k-means method, whereas \cite{lamirel2012new} used the neural clustering method. Afterward, the association and mapping among the clusters at different time is examined based on several heuristics. For example, \cite{oliveira2010mec} used cluster centroid related statistics, called comprehensive representation of clusters. This approach is very similar to the notion of detecting recurrent concept drifts in a semi-supervised context, see \cite{li2012mining} for an example.
A different method, called label-based diachronic approach \citep{lamirel2012new}, exploits the MultiView Data Analysis paradigm among the cluster labels at different time. In this approach, each feature is analyzed individually to compute recall, precision and F-measure. These information are used to construct heuristics for monitoring evolution. Our approach is different than the above methods, because: (a) we do not aim to propose a cluster monitoring method explicitly and (b) we do not use a static clustering method.
Besides the above methods, \cite{ferlez2008monitoring} proposed a joint clustering-monitoring method which uses the cross association algorithm to cluster data and a bipartite graph to monitor evolution. For data clustering, they group the distinct features (word) in each cluster and hence features do not coexist in different clusters. This is different than us as we exploit all the features in order to provide a feature level interpretation for the evolution.

The task of evolution interpretation aims to explain the reason for the evolution of clusters at different time. It can be accomplished by explicitly analyzing the features. To this aim, \cite{lamirel2012new} used the F-measures from individual features of the matched clusters (of different time) and construct a similarity report. In our work, this interpretation can be directly obtained from the link parameters by applying threshold on the link parameters values. Therefore, our method is different from \cite{lamirel2012new} as the link parameters computation is an integral part of the clustering task.

Based on the above distinctions from several viewpoints (clustering, monitoring and interpretation), we find that our method is more similar to the evolutionary clustering methods rather than the evolution monitoring methods. Therefore, we compare our method only with the relevant state-of-the-art evolutionary clustering methods, such as \cite{xu2014adaptive}, \cite{blei2006dynamic} and \cite{kim2015temporal}.

Now we focus on the literature related to our proposal. The idea of parametric link in a transfer learning context \citep{beninel2012parametric} is inherited from the concept for Generalized Discriminant Analysis (GDA) \citep{biernacki2002generalized}. GDA adapts the classification rule from a source population to a target population through a linear link map of their descriptive parameters. This is different than standard discriminant rules which assumes a similarity between the source and target populations. \cite{biernacki2002generalized} proposed several models with associated estimated parameters for GDA within the context of multivariate Gaussian distribution. Later, \cite{jacques2010extension} extends the work of \cite{biernacki2002generalized} for binary data using Bernoulli distribution \citep{bishop2006pattern}. We observe that these approaches can be exploited for developing an evolutionary clustering method by replacing the notion of source/target with different time epochs $t-1$/$t$. Besides, such development requires the derivation of the linear link for the Multinomial distribution. Afterward, the link parameters naturally allow us to interpret the evolution of the clusters at different time. 

Categorical data/observations consists of the responses from a certain number of categories. Different types (nominal and ordinal) of categorical data are observed in numerous studies \citep{agresti2002categorical}, such as social science, biomedical science, genetics, education and marketing. 
Moreover, data from different tasks, such as text retrieval and visual object classification, are often converted to the categorical form. For example, text data can be converted to this form by considering the unique words of the vocabulary as an independent category/term and then each sentence/paragraph/document is represented as a discrete count vector \citep{zhong2005generative}. 
The Multinomial distribution is a standard probability distribution for modeling and analyzing the discrete categorical data \citep{agresti2002categorical}. 

The Multinomial Mixture (MM) is a statistical model based on the Multinomial distribution. It has been used for cluster analysis with discrete data \citep{meilua2001experimental, agresti2002categorical, zhong2005generative, silvestre2014identifying, hasnat2015comparative}. \cite{meilua2001experimental} studied several Model-Based Clustering (MBC) methods with MM and experimentally compared them using different criteria such as clustering accuracy, computation time and number of selected clusters. \cite{silvestre2014identifying} proposed a MBC method for MM which integrates both model estimation and  selection task within a single EM algorithm. In their work, they extended the MBC strategy previously proposed by \cite{figueiredo2002unsupervised} and provided a formulation to compute the Minimum Message Length (MML) criterion for model selection. Most recently, \cite{hasnat2015comparative} proposed a MBC method which performs simultaneous clustering and model selection using the MM. Their strategy performs similar task as \cite{silvestre2014identifying} in a computationally efficient manner which has been previously proposed for the Gaussian distribution \citep{garcia2010simplification} and Fisher distribution \citep{hasnat2015mbhcfmm}. Moreover, similar to \cite{meilua2001experimental}, they provided a comparison among different model initialization and selection strategies. Following all of the above approaches \citep{meilua2001experimental, silvestre2014identifying, hasnat2015comparative}, in this research we exploit the MBC framework to cluster discrete data with MM.

MBC \citep{fraley2002model, melnykov2010finite} is a well-established method for cluster analysis and unsupervised learning. It assumes a probabilistic model (e.g., mixture model) for the data, estimates the model parameters by optimizing an objective function (e.g., model likelihood) and produces probabilistic clustering. The Expectation Maximization (EM) \citep{mclachlan2008em} is mostly used in MBC to estimate the model parameters. EM consists of an Expectation step (E-step) and a Maximization step (M-step) which are iteratively employed to maximize the log likelihood of the data.

Initialization of the EM algorithm has significant impact on clustering results \citep{mclachlan2008em, baudry2015mixtures}. The EM algorithm is sensitive to its initialization, because with different initializations it may converge to different values of likelihood function, some of which can be local maxima (i.e., sub-optimal results). In order to overcome this, numerous different initialization strategies are proposed and experimented in the relevant literature \citep{biernacki2003choosing, meilua2001experimental, baudry2015mixtures, hasnat2015comparative}. Following recommendations, we use the small-EM \citep{biernacki2003choosing, biernacki2006model, baudry2015mixtures, hasnat2015comparative} method to initialize the MM parameters.

MBC has been commonly exploited to identify the best model for the data by fitting a set of models with different parameterizations and/or number of components and then applying a statistical model selection criterion \citep{fraley2002model,biernacki2000assessing,figueiredo2002unsupervised,melnykov2010finite,hasnat2015mbhcfmm}. In this paper, we apply this model fitting and selection strategy for two purposes: (a) to identify the parametric submodels (Section \ref{ssec:param_sub_models}) and (b) to automatically select the number of components (Section \ref{ssec:varying_num_cluster}).
%
%
\subsection{Statistical model for evolutionary data samples}
\label{ssec:model_data}
Let $S^t$ be a set of samples corresponding to time $t$ and $S^{t+1}$ be a set from the next time $t+1$. We assume that while the cluster labels for $S^t$ are known to us (estimated from $t-1$), labels of $S^{t+1}$ are unknown.

Let $S^t$ be composed of $N^t$ pairs $(\bx_1^t, \bz_1^t),  \ldots , (\bx_{N^t}^t, \bz_{N^t}^t)$ where $\bx_i^t= \big\{ x_{i,1}^t, \AB \ldots, \AB x_{i,D}^t \big\}$ is the $D$ dimensional count vector of order $V$, i.e., $\sum_{d=1}^{D}x_{i,d}^t =V$ and $\bz_i$ is the associated class label such that $\bz_{i,k}^t=1$ if the data belongs to cluster $k$ with $k=1, \ldots ,K$ and $\bz_{i,k}^t=0$ otherwise. We assume that any sample $\bx_i^t$ of $S^t$ is an independent realization of the random variable $\bX^t$  of distribution:
\begin{equation*}
\bX^t  \sim \mathcal{M}(V, \bmu_k^t), \,\,\, k=1, \ldots ,K
\end{equation*}
with $\mathcal{M}(V, \bmu_k^t)$ is the $V$-order Multinomial distribution with parameter $\bmu_k^t=(\mu_{k,1}^t,\ldots,\mu_{k,D}^t)$ which is formally defined as \citep{bishop2006pattern}: 
\begin{equation}
\label{eq:mult_dist}
\mathcal{M}(\bx_i |V,  \bmu_k) = \begin{pmatrix}
 V \\ 
 x_{i,1}, x_{i,2}, \ldots, x_{i,D} 
\end{pmatrix} \prod_{d=1}^{D} \mu_{k,d}^{x_{i,d}}
\end{equation}
here, $\bmu_{k}$ is the parameter of the Multinomial distribution of $k^{th}$ class with $0 \leq \mu_{k,d} \leq 1$ and $\sum_{d=1}^{D}\mu_{k,d}=1$. Therefore, samples of the entire set $S^t$ can be modeled with a mixture of $k$ Multinomials, also called Multinomial Mixture (MM) model, which has the following form:
\begin{equation}
\label{eq:mm}
f\left ( \bx_i|\Theta_{K} \right ) = \sum_{k=1}^{K}\pi_{k} \, \mathcal{M}(\bx_i | V,  \bmu_k) 
\end{equation}

In Eq. (\ref{eq:mm}), $\Theta_{K} = \left \{ (\pi_1,\bmu_1), \ldots , (\pi_K,\bmu_K)\right \}$ is the set of model parameters, $\pi_k$ is the mixing proportion with $\sum_{k=1}^{K}\pi_{k}=1$ and $\mathcal{M}(\bx_i | V, \bmu_k)$ is the density function (Eq. (\ref{eq:mult_dist})). Besides, we assume that the class label $\bz_i^t$ is an independent realization of a random vector $\bZ^t$, distributed according to 1-order Multinomial:
\begin{equation*}
\label{eq:label_dist_mult}
\bZ^t\sim \mathcal{M}(1, \bpi^t)
\end{equation*}
where $\bpi^t = \pi_1^t,\ldots,\pi_K^t$ is the mixing proportion of the model in Eq. (\ref{eq:mm}).

The assumption of MM is similar for the samples of $S^{t+1}$ with random variable $\bX^{t+1} $ and parameter $\bmu_k^{t+1}$. However, for $S^{t+1}$ the labels $\bz_i^{t+1}$ of $N^{t+1}$ pairs $(\bx_1^{t+1},\bz_1^{t+1}), \ldots, (\bx_{N^{t+1}}^{t+1}, \bz_{N^{t+1}}^{t+1})$ are unknown. In the context of evolutionary clustering, our goal is to estimate the unknown labels $\bz_i^{t+1}$  for $i=1,\ldots,N^{t+1}$ using the information from $S^t$ and $S^{t+1}$ by establishing a link between $\bmu_k^t$ and $\bmu_k^{t+1}$.
%

\subsection{Parametric link/relationship among temporal data}
\label{ssec:param_link}
 
For random variables $Y^t$ and $Y^{t+1} $ distributed according to the Gaussian distribution, a linear distributional link exists (under weak assumptions) \citep{biernacki2002generalized}, which has the form: $Y^{t+1} \sim D Y^t+ b$, where $D$ and $b$ are the link parameters among the samples of different time epoch. 
For binary data the following distributional linear link among Bernoulli parameters ($\alpha^{t+1}$ and $\alpha^t$ with $0 \leq \alpha \leq1$) is derived by \cite{jacques2010extension}:
\begin{equation}
\label{eq:link_bern}
\alpha^{t+1}= \Phi \left ( \delta \, \Phi^{-1} \left ( \alpha^t \right ) + \lambda \, \gamma  \right )
\end{equation}
where $\delta \in \mathbb{R}^{+} \char`\\ \lbrace 0 \rbrace$, $\lambda \in \lbrace -1,1\rbrace$ and $\gamma \in \mathbb{R}$ are the link parameters. $\Phi$ is the cumulative Gaussian function of mean 0 and variance 1, see Fig. \ref{fig:cum_func}. We can use the above formulation for Multinomial parameters by considering two issues: (1) Multinomial parameter $\bmu_k$ has equivalent property as $\balpha_k$ except $\sum_{d=1}^{D} \mu_{k,d} = 1$ and (2) samples from $X$ are not necessary to be binary, which makes $\lambda$ useless. Considering these issues we can derive parametric link between $\bmu_t$ and $\bmu_{t+1}$ as: 
\begin{equation}
\label{eq:link_mult}
\mu_{k,d}^{t+1}=  \frac{\Phi \left ( \delta_{k,d} \,\, \Phi^{-1} \left ( \mu_{k,d}^t \right ) + \, \gamma_{k,d}  \right )}{\sum_{r=1}^{D} \Phi \left ( \delta_{k,r} \,\, \Phi^{-1} \left ( \mu_{k,r}^t \right ) + \, \gamma_{k,r}  \right )}  
\end{equation}
where $\delta_{k,d} \in \mathbb{R}^{+} \char`\\ \lbrace 0 \rbrace$ and $\gamma_{k,d} \in \mathbb{R}$ are the link parameters.
In Eq. (\ref{eq:link_mult}), the combination of parameters $\delta_{k,d}$ and $\gamma_{k,d}$ for $\forall k,d$ is called a full model which is over-parameterized and may leads to ambiguity. Instead, we consider several sub-models with certain constraints on the parameters, see the following section.
\subsection{Parametric sub-models}
\label{ssec:param_sub_models}
The idea of defining sub-models is frequent in Model-Based Clustering (MBC) \citep{fraley2002model}. We fit the evolutionary clustering model (Eq. (\ref{eq:link_mult})) with different sub-models and then select the best model using the Bayesian Information Criteria \citep{schwarz1978estimating}:
\begin{equation}
\label{eq:bic}
BIC= -2 L(\Theta) + \nu log \left ( N^{t+1} \right )
\end{equation}
where $L(\Theta)$ is the log-likelihood (Eq. (\ref{eq:llh_comp})) value associated to the MM parameters of $t+1$, $\nu$ is the number of free parameters of the sub-model. These sub-models provide sufficient interpretation about the change in parameters from time $t$ to $t+1$. Definition and interpretation of several basic sub-models, defined as pair ($\delta_{k,d}/\gamma_{k,d}$) are given below:
%

\textbf{(M1) $1/0$:} This model is constrained with $\delta_{k,d}=1$ and $\gamma_{k,d}=0$ for $\forall k,d$, i.e., $\nu=0$. It indicates that the observations $X^{t+1}$ can be modeled with $\mu_{k,d}^t$ and hence no evolution occurred. 

\textbf{(M2) $0/\gamma_{k,d}$:} This model is constrained with $\delta_{k,d}=0$ for $\forall k,d$, i.e., $\nu=K*D$. It indicates that the observations $X^{t+1}$ should be modeled without considering $\mu_{k,d}^t$. This model should be selected when a new cluster evolved independently and does not consider any historical information. This is the most general model that can certainly fit the observations $X^{t+1}$ to a MM most efficiently subject to a good initialization of the alternative iterative method. Several possible variations\footnote{\label{fnote1}Subscript $k$ means cluster dependent and $d$ means feature dependent. No subscription means a constant value for all clusters and features.} of this model are: $0/\gamma$, $0/\bgamma_k$ and $0/\bgamma_d$.

\textbf{(M3) $\delta_{k,d}/0$:} This model is constrained with $\gamma_{k,d}=0$ for $\forall k,d$, i.e., $\nu=K*D$. 
It indicates that $\mu_{k,d}^{t+1}$ are evolved through $\mu_{k,d}^t$ in a specific transformation space (inversed cumulative Gaussian). 
This model should be selected when true evolution occurred which can be explained in detail through certain belief on observed features and obtained clusters. Moreover, such a model can be plugged in with any other method in order to describe the cluster evolution. Several possible variations
of this model are: $\delta/0$, $\bdelta_k/0$ and $\bdelta_d/0$. This model is equivalent to the fundamental unconstrained model assumed by \cite{biernacki2002generalized}.

\textbf{(M4) $1/\gamma_{k,d}$:} In this model, $\delta_{k,d}=1$ for $\forall k,d$, i.e., $\nu=K*D$. This model does nearly similar task as model M3. It is relatively easier to fit through the additive term in the inverse cumulative Gaussian space. On the other hand, it is less expressive in terms of interpretation. Several possible variations
of this model are: $1/\gamma$, $1/\bgamma_k$ and $1/\bgamma_d$.
 
\subsection{Parameter estimation}
\label{ssec:param_est}
In our proposed formulation of evolutionary  clustering, we estimate two different types of parameters (see Eq. (\ref{eq:link_mult})): (1)  MM model parameters: $\mu$ and $\pi$ and (2) temporal link parameters: $\delta$ and $\gamma$. We estimate them in two steps. The first step consists of estimating $\mu$ and $\pi$ (only for $t=1$) for the observed samples of time $t$. In the second step, we estimate $\delta$ and $\gamma$. At any time epoch, we estimate the class labels $\bz_i$ by \textit{maximum a posteriori}.

\subsubsection{Multinomial Mixture (MM) Parameters}
\label{ssec:mm_param_est}
At time $t=1$, we estimate the MM parameters using an Expectation Maximization (EM) algorithm that maximizes the log-likelihood value which has the following form:
\begin{equation}
\label{eq:llh_comp}
L\left ( \Theta \right ) = \sum_{i=1}^{N} log \sum_{j=1}^{K} \pi_j \mathcal{M}\left ( \bx_i|\bmu_j \right )
\end{equation}
where $N= N^{1}$ is the number of samples. In the Expectation step (E-step), we compute posterior probability as:
\begin{equation}
\label{eq:posterior_cl}
\rho_{i,k} = p\left (z_{i,k}=1|\mathbf{x}_i \right ) = \frac{\pi_k \, \prod_{d=1}^{D}\mu_{k,d}^{x_{i,d}}}
{\sum_{l=1}^{K}\pi_{l} \, \prod_{d=1}^{D}\mu_{l,d}^{x_{i,d}}}
\end{equation}
In the Maximization step (M-step), we update $\pi_k$ and $\mu_{k,d}$ as:
\begin{equation}
\label{eq:maximization_cl}
\pi_k=\frac{1}{N}\sum_{i=1}^{N}\rho_{i,k}
\;\;\;
\text{and}
\;\;\; 
\mu_{k, d}=\frac{\sum_{i=1}^{N}\rho_{i,k} \,\mathbf{x}_{i, d}}{\sum_{i=1}^{N} \sum_{r=1}^{D} \rho_{i,k} \,\mathbf{x}_{i, r}}
\end{equation}
The E and M steps are iteratively employed until certain convergence criterion (difference of the log-likelihood values of successive iterations) is satisfied. The estimation of $\mu_{k,d}$ using Eq. (\ref{eq:maximization_cl}) is only applicable for $t=1$ due to the unavailability of any temporal information.  For any time $t+1$, when the link parameters are available, $\mu_{k,d}$ is estimated with Eq. (\ref{eq:link_mult}). 
\subsubsection{Link parameters}
\label{sssec:link_param_est}
Estimation of link parameters $\delta_{k,d}$ and $\gamma_{k,d}$ uses $\mu_{k,d}^t$ and the observed samples at time $t+1$. Similar to \cite{jacques2010extension}, we use again an EM algorithm, but in which the M step is not explicit. Consequently, we employ an external optimization method such as an alternative iterative algorithm which consists of a succession, componentwise of the simplex method\footnote{For the implementation, we used \textbf{\textit{neldermead}} function of \textbf{\textit{nloptr}} R package \citep{ypma2014introduction}. The lower and upper bounds were set to $-2.5$ and $+2.5$ respectively only for the $\gamma_{k,d}$ parameters.} \citep{nelder1965simplex}. 
In general, the starting point of the alternative algorithm corresponds to the case when $\mu_{k,d}^{t+1} = \mu_{k,d}^t$, i.e.,  $\delta_{k,d} =1$ and $\gamma_{k,d}=0$. However, in order to obtain a better estimate and save computation time \footnote{The simplex method requires a large number of iterations to converge.}, we apply an efficient approach, see Section \ref{sssec:link_param_init}.

\begin{algorithm}[h]
 \SetAlgoLined
 \KwIn{$\chi= \left \{ S^{t}  \right \}_{t=1,\ldots,T} ,\,   S^{t} = \left \{  \bx_i \right  \}_{i=1, \ldots, N^t}, \, \bx_i = \left \{  x_{i,d} \right  \}_{d=1, \ldots, D}, \, x_{i,d} \in \mathbb{N}$}
 \KwOut{Evolutionary clustering of $\chi$ with $K$ classes and link parameters: $\delta_{k,d}^t$ and $\gamma_{k,d}^t$ $\forall k,d,t$.}
 \ForEach{ $t$}{
 \If{$t=1$}
 {
    Initialize $\pi_{j,k}$ and $\mu_{j,k}$ for $1\leq j\leq k$ using the \emph{small-EM} procedure, see Section \ref{ssec:mm_param_init}\;
 }
 \While{not converged}{
  \{Perform the E-step of EM\}\;
  \ForEach{ $i$ and $j$}{
  Compute $\rho_{ik}=p(z_{i,k}=1|\bx_i)$ using Eq. (\ref{eq:posterior_cl})
  }
  \{Perform the M-step of EM\}\;
  \For{$k=1$ to $K$}{
  \uIf{$t=1$}
 {
  Update $\pi_k$ and $\bmu_k$ using Eq. (\ref{eq:maximization_cl})
  }
  \Else
{
    Update $\pi_k$ using Eq. (\ref{eq:maximization_cl}) \\
    Compute $\delta_{k,d}$ and $\gamma_{k,d}$, see Sec. \ref{sssec:link_param_est} \\
    Update $\bmu_k$ using Eq. (\ref{eq:link_mult})
}   
  }
  } 		
}
 \caption{Algorithm for clustering using parametric link among multinomial mixtures (PLMM).}
 \label{algo:pl_ecl}
\end{algorithm}
\subsection{Parameters initialization}
\label{ssec:param_init}
In the proposed clustering method (Algorithm \ref{algo:pl_ecl}), we need to initialize both the MM parameters $\Theta^{init}_K = \big \{ (\pi^{init}_1,\bmu^{init}_1), \AB \ldots \AB , (\pi^{init}_K,\bmu^{init}_K) \big\}$ for time $t1$ and the link parameters ($\delta$ and $\gamma$). 

\subsubsection{Multinomial Mixture (MM) Parameters}
\label{ssec:mm_param_init}
Generally, the MM parameters are initialized randomly \citep{meilua2001experimental, hasnat2015comparative}. However, with both synthetic and real data it has been demonstrated by \cite{hasnat2015comparative} that, random initialization has its limitation w.r.t. the clustering performance and stability. Therefore, following \cite{hasnat2015comparative}, we initialize the model parameters using the small-EM procedure. This small-EM procedure consists of running multiple short runs of randomly initialized EM and then selecting the one with the maximum likelihood value. Here, short run means that the EM procedure does not need to wait until convergence and it can be stopped when a certain number of iterations is completed.

\subsubsection{Link parameters}
\label{sssec:link_param_init}
We propose an initialization procedure based on the predictive parameters set for next time epoch $\Theta^{pred}_K = \big \{ (\pi^{pred}_1,\bmu^{pred}_1), \AB \ldots \AB , (\pi^{pred}_K,\bmu^{pred}_K) \big \}$ . Let $\Theta^{t}_K=\left \{ (\pi^{t}_1,\bmu^{t}_1), \ldots , (\pi^{t}_K,\bmu^{t}_K)\right \}$ is the set of parameters for the current time ($t$) epoch. Our initialization procedure consists of the following steps:
\begin{itemize}
\item Step 1: estimate $\Theta^{pred}_K$ using data samples of next time $X^{t+1}$
and an EM algorithm which is initialized with $\Theta^{t}_K$.
\item Step 2: compute $\delta^{init}_{k,d}$ and $\gamma^{init}_{k,d}$ for each $k$ and $d$ as:
\begin{equation}
\label{eq:init_m2}
\gamma^{init}_{k,d} = \Phi^{-1} \left ( \mu_{k,d}^{pred} \right ) \;\;\; \text{for model M2}
\end{equation}
\begin{equation}
\label{eq:init_m3}
\delta^{init}_{k,d} = \frac{\Phi^{-1} \left ( \mu_{k,d}^{pred} \right )}{\Phi^{-1} \left ( \mu_{k,d}^{t} \right )} \;\;\; \text{for model M3}
\end{equation}
\begin{equation}
\label{eq:init_m4}
\gamma^{init}_{k,d} = \Phi^{-1} \left ( \mu_{k,d}^{pred} \right ) - \Phi^{-1} \left ( \mu_{k,d}^{t} \right ) \;\;\; \text{for model M4} 
\end{equation}
\end{itemize}
The Eq. (\ref{eq:init_m2}), (\ref{eq:init_m3}) and (\ref{eq:init_m4}) are simply derived from Eq. (\ref{eq:link_mult}) with the consideration that denominator is equal to 1, i.e., $\sum_{d=1}^{D}\mu_{k,d}=1$ for $k=1, \ldots, K$.
\subsection{Varying number of clusters}
\label{ssec:varying_num_cluster}
The methodology presented in the previous sub-sections assumes the same number of clusters $K$ for each time epoch. In this sub section, we propose an extension of it such that the method can handle varying $K$ at different time, i.e., $K_{t}$ and $K_{t+1}$ may be different. To this aim, we modify the links initialization strategy (Section \ref{sssec:link_param_init}) in order to adapt the variability among $\Theta^{t}_{K_{t}}$ and $\Theta^{t+1}_{K_{t+1}}$. At time epoch $t$, this extended method requires additional information, such as: (a) number of clusters $K_{t+1}$ and 
(b) cluster mapping between $\Theta^{t}_{K_{t}}$ and $\Theta^{t+1}_{K_{t+1}}$. 

We adopted the method proposed by \cite{hasnat2015comparative} with L-method \citep{ salvador2004determining} to select the number of cluster automatically at each time epoch. In order to initialize the link parameters, first we select the number of clusters $K_{t+1}$ and obtain the predictive parameter set $\Theta^{pred}_{K_{t+1}}$.
Next, for each cluster $k$ in  $\Theta^{pred}_{K_{t+1}}$ we find the corresponding cluster in $\Theta^{t}_{K_{t}}$ based on the minimum symmetric kullback leibler divergence (sKLD). sKLD among two clusters $a$ and $b$ is defined as \citep{hasnat2015comparative}:
\begin{equation}
\label{eq:kld}
\begin{aligned}
sKLD = \frac{D_{KL}\left ( \bmu_a, \bmu_b \right ) + D_{KL}\left ( \bmu_b, \bmu_a \right )}{2}, \;\; \text{where} \\
D_{KL}\left ( \bmu_a, \bmu_b \right ) = \sum_{d=1}^{D} \mu_{a,d}\, ln\left ( \frac{\mu_{a,d}}{\mu_{b,d}} \right )
\end{aligned}
\end{equation}
After establishing the correspondences, we use Eq. (\ref{eq:init_m2}), (\ref{eq:init_m3}) and (\ref{eq:init_m4}) to set the initial values of the link parameters. Finally, we estimate the link parameters following Section \ref{sssec:link_param_est}.
\subsection{Interpretation of cluster evolution}
\label{ssec:interpretation_evolution}
The link parameters ($\delta_{k,d}$ and $\gamma_{k,d}$) along with the function $\Phi$ are the key to interpret the cluster evolution. Let us notice some basic interpretation of the values of these parameters for all feature $d$ and cluster $k$:
\begin{itemize}
\item $\delta_{k,d}=0$ means that $\mu_{k,d}$ (probability) at $t+1$ does not depend on $t$, whereas $\delta_{k,d}=1$ (with $\gamma_{k,d}=0$) means identical probability at two different times.
\item $\delta_{k,d}\rightarrow 0$ and/or $\gamma_{k,d}\rightarrow \infty$ means that the distribution \textit{tends to uniform} distribution.
\item $\delta_{k,d}\rightarrow \infty$ and/or $\gamma_{k,d}\rightarrow -\infty$ means that the distribution tends to be \textit{more concentrated} (Dirac distribution) at time $t+1$ in the feature which has the highest probability at time $t$.
\end{itemize}
\begin{figure}[h]
\centering
\includegraphics[scale=0.42]{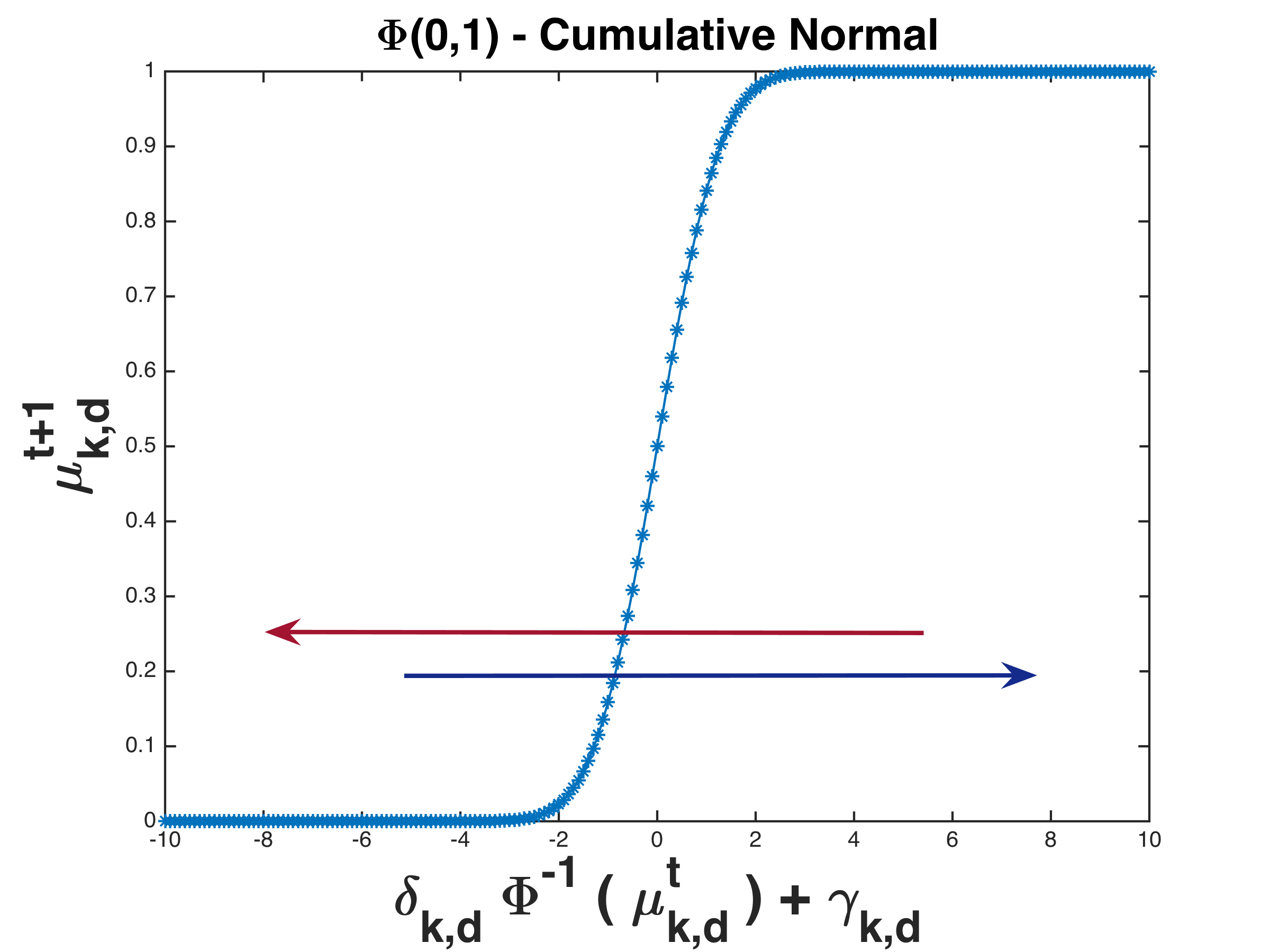} 
\caption{Illustrations of Cumulative Gaussian function and its relationship with the parameter change of Multinomial distribution using Eq. (\ref{eq:link_mult}). The arrows indicates the direction of changes in the inverse function space which eventually increase/decrease the probability.}
\label{fig:cum_func}
\end{figure}

In order to get further interpretation, we need to understand the Multinomial parameters $\mu_{k,d}$ and the space spanned by the cumulative Gaussian $\Phi$ and its inverse $\Phi^{-1}$. Let us consider an experiment of drawing $V$ balls of $d=1,\ldots,D$ different colors (represent features). After each draw, the color of the ball is recorded in a $D$ dimensional count vector $\bx_i$ and the ball is replaced. Therefore, at the end of $i^{th}$ experiment $\bx_{i,d}$ reveals the count of drawing the $d^{th}$ colored ball. When a Multinomial distribution is used to fit such experimental data, its parameter $\mu_{k,d}$ reveals the probability of drawing the $d^{th}$ colored ball.

Now, let us consider $\Phi$ in Fig. \ref{fig:cum_func} where the values along the Y-axis represent the possible values of  $\mu_{k,d}^{t+1}$ (with $0 \leq \mu_{k,d}^{t+1} \leq 1$) and the X-axis represents the values of  $\mu_{k,d}^{t}$  after transforming through $\Phi^{-1}$ function. Now, according to Eq. (\ref{eq:link_mult}), cluster evolutions ($\mu_{k,d}^{t} \; \rightarrow \; \mu_{k,d}^{t+1}$) can be explained through multiplication (using $\delta_{k,d}$) and addition/subtraction (using $\gamma_{k,d}$) operations.

The values of $\gamma_{k,d}$ can certainly indicates the increase/decrease of the probability of certain feature (color) subject to the selection of sub-model \textbf{M4}. On the other hand if sub-model \textbf{M3} is selected, values of $\delta_{k,d}$ can explain the belief that $\mu_{k,d}^{t+1}$ should decrease if $\mu_{k,d}^{t} < 0.5$ and increase if $\mu_{k,d}^{t} > 0.5$. For example, let us consider that in a 2 colors (red and green) ball experiment the probability of the red color ball is changed from 0.8 (at time t1) to 0.7 (at time t2). Such a change can be explained with model \textbf{M3} with $\delta_{k,red}=0.6$, which indicates that the belief is decreased at the next time. From the above discussions it is evident that the proposed method is capable to interpret the cluster evolutions up to the feature level.
%
\section{Numerical experiments}
\label{sec:exp_res}

We begin the experiments using simulated evolutionary data samples and evaluate w.r.t. the state-of-the-art methods. A characteristic comparison of different methods is presented in Table \ref{tab:methods_comp}. For the simulated samples; we use the Adjusted Rand Index (ARI) \citep{hubert1985comparing} as a measure for evaluation. Next, we experiment and compare methods using real data. 
We use one of the real datasets experimented by \cite{kim2015temporal}. 
We choose the \textit{political opinion dataset} from the ImagiWeb project \citep{velcin2014investigating} as it consists of data from an interesting time period - during and after the election.
%

\begin{table}[h]
\centering
\caption{Characteristic comparison of different state-of-the-art evolutionary clustering methods: Parametric Link among Multinomial Mixtures (PLMM, our proposed method), Temporal Multinomial Mixture (TMM) \citep{kim2015temporal}, Dynamic Topic Model (DTM) \citep{blei2006dynamic} and adaptive evolutionary clustering method (AFFECT) \citep{xu2014adaptive}.}
\begin{tabular}{|c|c|c|c|c|}
\hline
\textbf{}                    & \textbf{PLMM} & \textbf{DTM} & \textbf{TMM} & \textbf{AFFECT} \\ \hline
\textbf{Data Type}           & Discrete            & Discrete         & Discrete         & Continuous           \\ \hline
\textbf{Interpret Evolution} & Yes             & No           & No           & No              \\ \hline
\end{tabular}
\label{tab:methods_comp}
\end{table}
\subsection{Simulated Data Samples}
\label{ssec:simulated_samples}
Following standard sampling methods we generate different sets $\lbrace S^t \rbrace _ {t=1,\ldots,T}$ of simulated data for different time epochs. We draw a finite set of categorical samples (discrete count vectors) $S^t= \{\bx_{i}\}_{i,...,N^t}$ with different numbers (10, 20 and 40) of features (dimensions) $D$. These samples are issued from Multinomial Mixture (MM) models of $K=3$ classes. We consider two different sets of samples: 
\begin{itemize}
\item Samples with higher order of categorical count (\textit{\textbf{hos}}) with $V \sim 1.5*D$ with 3 time epochs each having different number of  i.i.d. samples: $N^1=500$, $N^2=100$, and $N^3=200$. We also add noisy counts with these samples. These type of samples provides better resemblance with the MM parameters due to sufficient number of count in the observations. Practically, this is similar to the fact when the observations consists of data over longer period of time.
\item Samples with lower order of categorical count (\textit{\textbf{los}}) with $V \sim 0.7*D$ with 5 time epochs each having different number of  i.i.d. samples: $N^1=50$, $N^2=40$, $N^3=40$, $N^4=30$ and $N^5=20$. This type of samples are sparse and often difficult to distinguish among clusters. Practically, this is similar to the fact when the observations consists of data over shorter period of time.
\end{itemize}

The evolutionary data generation process consists of two steps: (1) determine MM parameters $\mu_{k,d}$ at each time epoch $t=1,\ldots, T$ and (2) sample observations from the specified MM following assumption specified by \cite{blei2003latent}. For $t=1$, we sample $\mu_{k,d}$ from a Dirichlet distribution and verify (separation w.r.t. the other clusters parameters \citep{silvestre2014identifying}) it using the symmetric Kullback-Leibler Divergence value. For $t>1$,  we sample $\mu_{k,d}$ from $\mu_{k,d}^{t-1}$ using the MM link relationship defined in Eq. (\ref{eq:link_mult}). This ensures that we maintain the temporal smoothness property \citep{chakrabarti2006evolutionary, xu2014adaptive} of the evolutionary data samples. In order to use the link relationship, we use only model M4 for \textit{hos} data samples and randomly select a model among M1, M3 and M4 for \textit{los} data samples. Next, we set the associated link parameters ($\delta_{k,d}$ and $\gamma_{k,d}$) randomly within a pre-specified range of values. 

To sample observations, first we choose the order $V_k$ of each cluster. Our sampling procedure for each observation $i$ at each time $t$ follows the steps below:
\begin{itemize}
\item Choose a cluster $z_{i,k}=1$ as: $\bz_i \sim \mathcal{M}\left ( 1, \pi_1, \ldots, \pi_D \right ), \text{with,} \, \pi_d=\frac{1}{k}$.
\item Choose the order $\tau_i$ of Multinomial for the sample $\bx_i$ using Poisson distribution as: $\tau_i \sim \text{Poisson}\left ( V_{z_i} \right )$.
\item Draw sample $\bx_i$ using Multinomial distribution as: $\bx_i \sim \mathcal{M} \big ( \tau_i, \mu_{k,1}, \AB \ldots \AB, \mu_{k,D} \big )$.
\end{itemize}
\begin{table}[h]
\centering
\caption{Simulated data evaluation and comparison using Adjusted Rand Index (ARI) \citep{hubert1985comparing}. Methods: PLMM (proposed), Dynamic Topic Model (DTM), Temporal Multinomial Mixture (TMM) and AFFECT with k-means. Datasets consist of different types (\textit{hos} and \textit{los}) of samples with different numbers (10, 20 and 40) of features. \textbf{\textit{hos}}: higher order samples and \textbf{\textit{los}}: lower order samples. \textbf{Boldfaced} indicate the best result and \underline{underlined} numbers indicate second best. Values inside the parentheses provide the standard deviation of the ARI values.}
\begin{tabular}{|c|c|c|c|c|}
\hline
\textbf{}   & \textbf{PLMM} & \textbf{TMM}  & \textbf{DTM} & \textbf{AFFECT} \\ \hline
\textbf{10, \textit{hos}} & \textbf{0.91} (0.07)           & \underline{0.86} (0.11)       & 0.79 (0.14)       & 0.43 (0.12)          \\ \hline
\textbf{10, \textit{los}} &\underline{0.81}	(0.19) & \textbf{0.91} (0.1)	&\underline{0.81} (0.1) 	&0.34 (0.11)          \\ \hline
\textbf{20, \textit{hos}} & \textbf{0.96} (0.05)          & \underline{0.91}  (0.1)      & 0.81 (0.18)       & 0.37 (0.11)           \\ \hline
\textbf{20, \textit{los}} &0.90 (0.18) 	&\textbf{0.98} (0.04)	&\underline{0.95}	(0.11) &0.35 (0.09)           \\ \hline
\textbf{40, \textit{hos}} & \textbf{0.97} (0.05)          & \underline{0.92} (0.11)       & 0.48 (0.4)       & 0.33 (0.11)           \\ \hline
\textbf{40, \textit{los}} &\underline{0.93}	(0.16) & \textbf{0.97}	(0.05) & \textbf{0.97}	(0.1) &0.36 (0.1)          \\ \hline
\end{tabular}
\label{tab:syn_data_eval}
\end{table}

We applied our proposed Parametric Link among Multinomial Mixtures (PLMM, Algorithm \ref{algo:pl_ecl}) clustering method on these simulated data using the basic sub-models defined in Sec. \ref{ssec:param_sub_models}. Table \ref{tab:syn_data_eval} provides the results using the ARI \citep{hubert1985comparing} measure. Moreover, it provides a comparative evaluation w.r.t. other state-of-the-art methods (see comparison in Table \ref{tab:methods_comp}): (a) Temporal Multinomial Mixture (TMM) \citep{kim2015temporal} with smoothness parameter $\alpha=1$; (b) Dynamic Topic Model (DTM) \citep{blei2006dynamic} with hyper-parameter $\alpha=0.01$ and (c) Adaptive evolutionary clustering method (AFFECT\footnote{We experimented AFFECT with hierarchical and spectral clustering also. However, k-means provided the best results.}) \citep{xu2014adaptive} with k-means and Euclidean distance as a measure of similarity. We compute the average ARI of time $t=2,\ldots,T$ (at $t=1$ there is no evolution). 
Results in Table \ref{tab:syn_data_eval} w.r.t. ARI evaluation shows that: 
\begin{itemize}
\item PLMM (proposed) provides highest ARI for the \textbf{\textit{hos}} samples and TMM \citep{kim2015temporal} provides highest ARI for the \textbf{\textit{los}} samples. These results are not surprising as both PLMM and TMM methods are specialized methods to cluster samples which are drawn from Multinomial distributions.
\item DTM \citep{blei2006dynamic} provides better results for \textbf{\textit{los}} samples and higher dimensional data. This type of data is more likely to extract from text documents for which DTM was originally proposed.
\item AFFECT \citep{xu2014adaptive} performs poorly compares to others for both types of sample. This is expected because of the similarity measure used in AFFECT is appropriate for continuous data.
\end{itemize}

Next, we test statistical hypothesis among PLMM, TMM and DTM using \textit{two sample t-test} at the 5\% significance level. The null hypothesis is that - the data in two results comes from independent random samples from normal distributions with equal means and equal but unknown variances. Results show that for all \textit{hos} data the hypothesis is rejected with p-value$<$0.001. On the other hand, for the \textit{los} data it is rejected only for 10 dimensional samples among the pairs (PLMM, TMM) and (DTM, TMM) with p-value$<$0.0001.

Next, we analyze the evolution of the clusters in terms of selected sub-models. Table \ref{tab:syn_evolution} provides the rate of different selected models. We see that, for the \textit{\textbf{hos}} data samples the model M4 ($1/\gamma_{k,d}$) is mostly selected. On the other hand, for the \textit{\textbf{los}} data samples, different models M1: ($1/0$), M4: ($1/\gamma_{k,d}$) and M3: ($\delta_{k,d}/0$) are selected at certain rate. This observation confirms that PLMM successfully recovers the cluster evolutions with different models which were used to generate the simulated data. Interestingly, we observe that the model M2 ($1/\gamma_{k,d}$) is not selected which reflects the true fact that it was not considered to generate the simulated data samples. Now based on the selected model, we can provide further interpretation using $\delta_{k,d}$ and $\gamma_{k,d}$, see Sec. \ref{ssec:param_sub_models}.

\begin{table}[h]
\centering
\caption{Percentage of the selected models for the interpretation of evaluation. \textbf{\textit{hos}}: higher order (categorical count) samples and \textbf{\textit{los}}: lower order samples. \textbf{Boldfaced} indicate the highest rate.}
\begin{tabular}{|c|c|c|c|c|}
\hline
\textbf{} & \textbf{M1: ($1/0$)} & \textbf{M4: ($1/\gamma_{k,d}$)}   & \textbf{M3: ($\delta_{k,d}/0$)}   & \textbf{M2: ($0/\gamma_{k,d}$)} \\ \hline
\textbf{10, \textit{hos}}                & 0 \%        & \textbf{94 \%} & 6 \%          & 0 \%       \\ \hline
\textbf{10, \textit{los}}                & 15 \%        & 38 \%          & \textbf{47 \%} & 0 \%       \\ \hline
\textbf{20, \textit{hos}}                & 0 \%       & \textbf{92 \%} & 8 \%         & 0 \%       \\ \hline
\textbf{20, \textit{los}}                & 14 \%       & \textbf{43 \%} & \textbf{43 \%} & 0 \%       \\ \hline
\textbf{40, \textit{hos}}                & 0 \%       & \textbf{96 \%} & 4 \%         & 0 \%       \\ \hline
\textbf{40, \textit{los}}                & 4 \%       & 37 \%         & \textbf{59 \%} & 0 \%      \\ \hline
\end{tabular}
\label{tab:syn_evolution}
\end{table}

Finally, we conduct experiments with varying number of clusters $K$ at different time epoch. For this experiment, we use the same MM parameters which were used to generate the \textit{ hos} data samples. To ensure different $K$ at different epoch, we randomly select a pair of time epochs and remove a cluster from one of them. Then, we generate $N^t=N^{t+1}=1000$ synthetic data samples from them using the same procedure mentioned before.
Applying the extension of PLMM method (Section \ref{ssec:varying_num_cluster}) on these data provides the following results (ARI): 0.967 (0.09) for $d=10$,  0.988 (0.04) for $d=20$ and 0.986 (0.05) for $d=40$. These results confirms that our proposed extension can cluster the synthetic data with varying $K$ and provides reasonable accuracy.
\subsection{IW-POD dataset}
\label{sssec:comparison_real_data}
We consider three different methods, Dynamic Topic Model (DTM) \citep{blei2006dynamic}, Temporal Multinomial Mixture (TMM) \citep{kim2015temporal} and Parametric Link among Multinomial Mixtures (PLMM), for a comparative evaluation of the performance on IW-POD dataset. These methods are selected based on their specialty to cluster discrete evolutionary/temporal data. We set 100 maximum number of iterations as the convergence criterion for all methods. Besides, we set the threshold log-likelihood difference values as 0.0001 for PLMM and TMM. The smoothness parameter $\alpha$ of TMM was set to 1. The DTM hyper-parameter $\alpha$ was set to 0.01. For the PLMM method, we consider the sub-models mentioned in Sec. \ref{ssec:param_sub_models}. 
%

%
\begin{figure}
\centering
\includegraphics[scale=0.32]{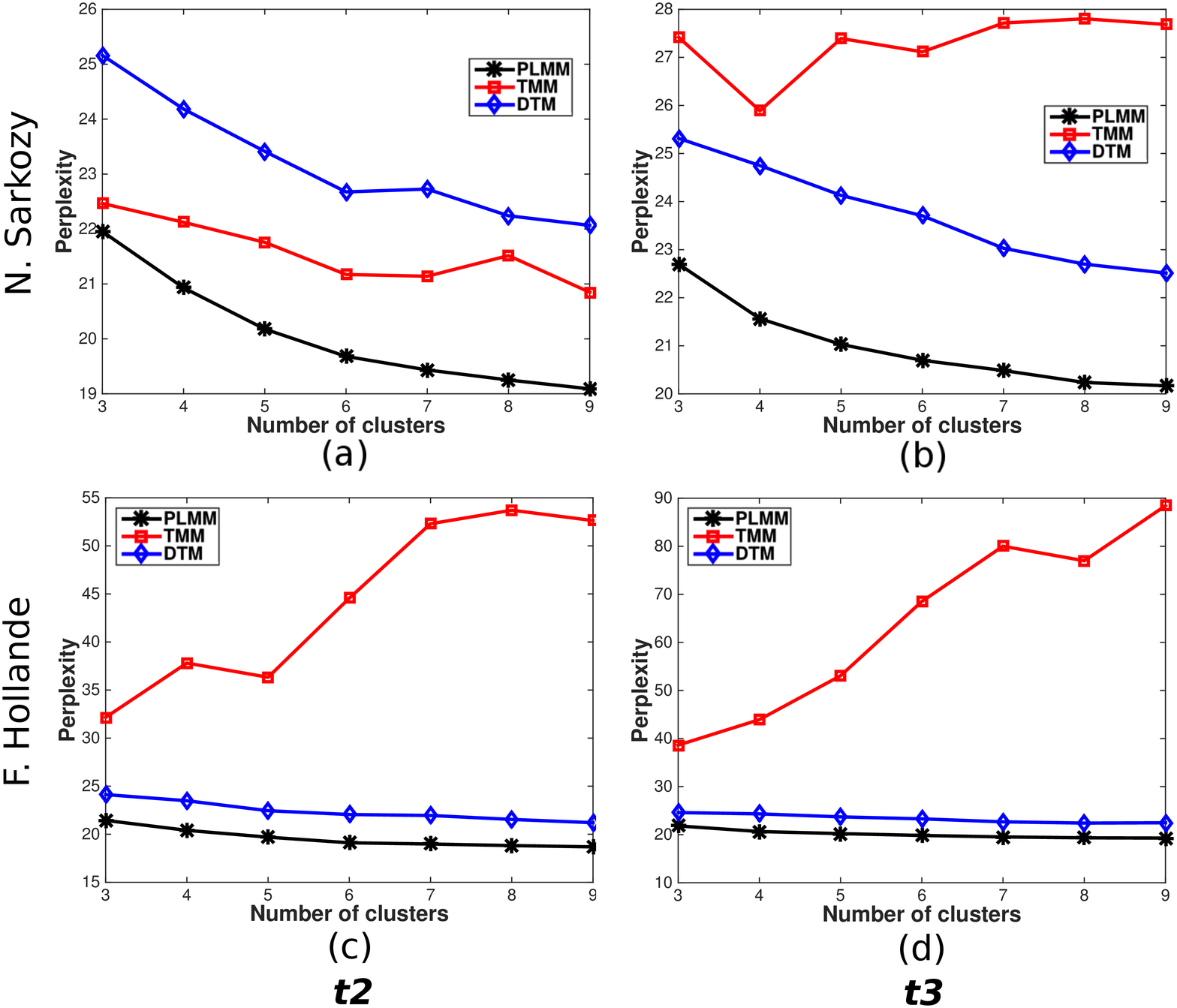} 
\caption{ Comparison of different methods w.r.t. the \emph{perplexity} values (\textbf{\textit{lower is better}}) computed from the IW-POD data of two entities (row-1: Sarkozy and row-2: Hollande) and two time epochs (column-1: epoch $t2$ and column-2: epoch $t3$). Methods: Dynamic Topic Model (DTM) \citep{blei2006dynamic}, Temporal Multinomial Mixture (TMM) \citep{kim2015temporal} and our proposed Parametric Link among Multinomial Mixtures (PLMM) method.}
\label{fig:real_perplexity}
\end{figure}
IW-POD dataset does not provide ground truth cluster labels, due to which we were unable to evaluate clustering results with the known-labels based metric such as \textbf{\textit{ARI}}. In this context, we evaluate the methods using a well known likelihood related measure called \emph{perplexity} on a held-out test set \citep{murphy2012machine, blei2003latent}. \textbf{\textit{Perplexity}} is a quantity originally used in the field of language modeling  \citep{murphy2012machine}. It measures how well a model has captured the underlying distribution of language. In clustering context, \emph{perplexity} is defined as the reciprocal geometric mean of the per feature (word) log-likelihood of a test set, which is computed using the model parameters learned with a training set. Therefore, the \emph{lower} \emph{perplexity} value indicates that the estimated (trained) model performs \emph{better} to fit the test data. \textbf{\textit{Perplexity}} can be formally defined as \citep{ blei2003latent}:
\begin{equation}
\label{eq:perplexity}
perplexity(X^{test}) = exp \left (  - \frac{L \left ( \Theta^{train} \right )}{\sum_{i=1}^{N^{test}} V_i } \right )
\end{equation}
where, $V_i$ is the total number of feature counts (words for document) in observation $i$, $L \left ( \Theta^{train} \right )$ denotes the log-likelihood of the test data set computed using the trained model parameters $\Theta^{train}$ and Eq. (\ref{eq:llh_comp}).

In our experiments, for each time epoch $t$, we compute \emph{perplexity} from 5 folds of training-test data division and then take the average of 5 \emph{perplexity} values as the final measure. For each fold, we used 80\% data for training the model and obtain parameters $\Theta^{train}$ and the remaining 20\% data to compute \emph{perplexity} using Eq. (\ref{eq:perplexity}). Fig. \ref{fig:real_perplexity} illustrates the perplexity values computed from the data of two entities (row-1: Sarkozy and row-2: Hollande) and two time epochs (column-1: epoch $t2$ and column-2: epoch $t3$). Time epoch $t1$ is not considered because it does not reflect the link relationship and temporal aspect of data clustering. 

Results in Fig. \ref{fig:real_perplexity} show that, PLMM provides the best \emph{perplexity} compared to DTM and TMM. This means that, compared to other methods, PLMM provides better fitting of the underlying Multinomial distribution to the test data. The next best (3 out of 4) method is the DTM followed by the TMM. Indeed, the results from TMM are intuitive as the fitted models are highly influenced by the other cluster components (Multinomial distributions) from the previous and next time epochs. In contrary, PLMM only consider the link from one cluster in the previous time epoch and fit the data accordingly.

\begin{figure}
\centering
\includegraphics[scale=0.5]{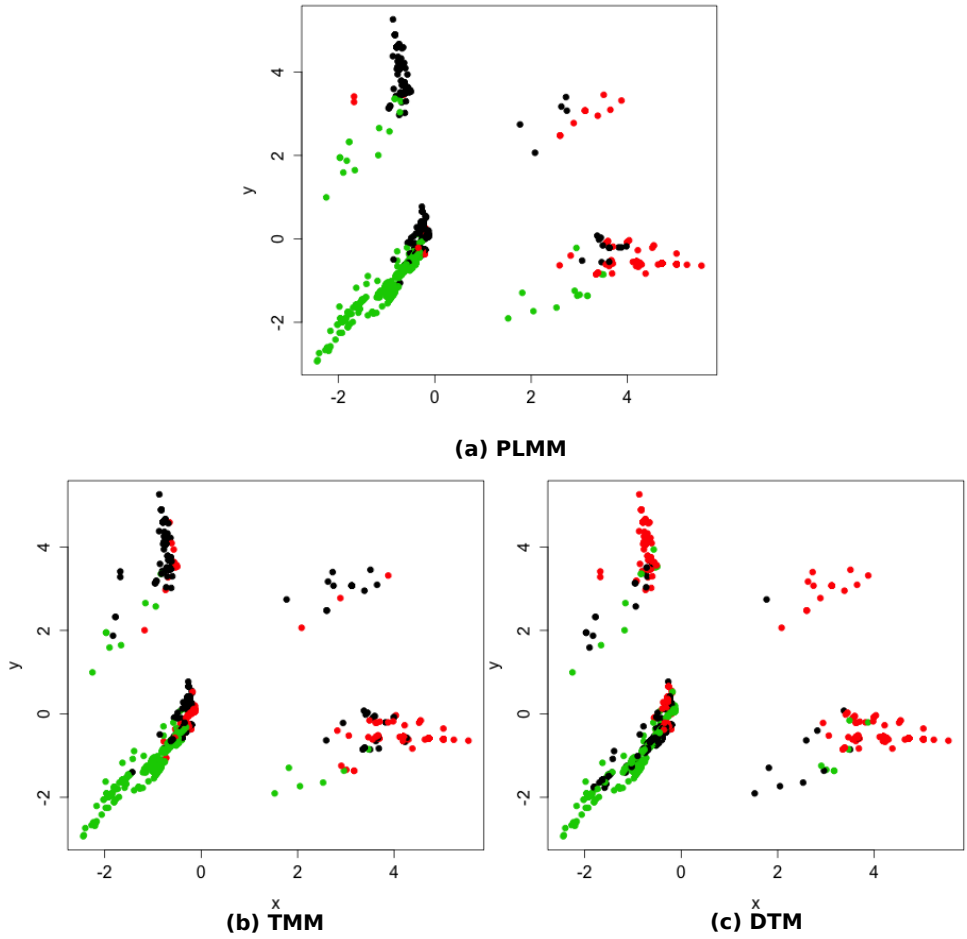} 
\caption{Illustration of clustering results visualized with Multidimensional scaling \citep{kruskal1978multidimensional}. Methods: (a) proposed Parametric Link among Multinomial Mixtures (PLMM); (b) Temporal Multinomial Mixture (TMM) \citep{kim2015temporal} and (c) Dynamic Topic Model (DTM) \citep{blei2006dynamic}.
}
\label{fig:visual_cl_res_comp}
\end{figure}

Fig. \ref{fig:visual_cl_res_comp} provides a visual illustration of clustering results obtained from the above three methods. This illustration is obtained by using the Multidimensional scaling \citep{kruskal1978multidimensional} technique where the distance matrix among the observations is computed by first converting the count vectors into probabilities and then using the sKLD (Eq. \ref{eq:kld}) as a measure of distance. The clustering results are obtained with $K=3$, time epoch $t2$ and the observations associated with the entity NS. From visual comparison among the plots in Fig. \ref{fig:visual_cl_res_comp}, we can say that PLMM provides better separation than TMM and DTM. Indeed, this observation agrees with the numerical results obtained with the \emph{perplexity} values in Fig. \ref{fig:real_perplexity}(a) for $K=3$.  

Next, we apply the extension of PLMM method (Section \ref{ssec:varying_num_cluster}) with this dataset and observe the \emph{perplexity} for time epochs $t2$ and $t3$. For the entity NS, we obtain average \emph{perplexity} values as: $t2:\,26.56$ and $t3:\,25.06$ where average $K_{t2}$ is 3 and average $K_{t3}$ is 5. For the entity FH, we obtain average \emph{perplexity} values as: $t2:\,13.08$ and $t3:\,5.17$ where average $K_{t2}$ is 4 and average $K_{t3}$ is 5. Compared to the results in Fig. \ref{fig:real_perplexity} we see that, \emph{perplexity} values increases (performance decreases) for entity NS and decreases (performance improves) for FH. Based on these observations, we can say that the extension of PLMM provides a good compromise in performance and works well for varying $K$ at different epochs. We do not compare these results with the TMM and DTM methods as they work with fixed $K$ for all time epochs.

Finally, let us focus on the interpretations of cluster evolutions in the IW-POD dataset. Table \ref{tab:details_evolution_iw} provides the selection rate of different models at different time epochs (see Table \ref{tab:iw_pod_details} for details of time division). Listed rates provide us very interesting observations from which we can say that:
\begin{itemize}
\item The opinions about NS were evolving almost similar way during and after the election period. These evolutions can be interpreted through the belief on aspects using models M3:($\delta_{k,d}/0$) (93\%) and M4:($1/\gamma_{k,d}$) (7\%). This indicates that during $t1$-$t2$-$t3$ opinions about NS were changing slowly.

\item Model M2:($0/\gamma_{k,d}$) is selected for all clusters of opinions about FH during t1-t2. This means that the opinions change significantly between $t1$ and $t2$ period.
From $t2$ to $t3$ (both after election period), opinions were evolving, which can be interpreted through the belief on the features with the models M4:($1/\gamma_{k,d}$) (62\%) and M3:($\delta_{k,d}/0$) (38\%).
\end{itemize} 
\begin{table}[h]
\centering
\caption{Selection rate of different models (Sec. \ref{ssec:param_sub_models}) for the IW-POD dataset at different time epochs (see Table \ref{tab:iw_pod_details} for details of time division).}
\begin{tabular}{|c|c|c|c|c|}
\hline
\textbf{}                    & \textbf{M1: ($1/0$)} & \textbf{M4: ($1/\gamma_{k,d}$)}   & \textbf{M3: ($\delta_{k,d}/0$)}   & \textbf{M2: ($0/\gamma_{k,d}$)} \\ \hline
\textit{\textbf{NS (t1-t2)}} & 0 \%       & 0 \%       & \textbf{100 \%}       & 0 \%       \\ \hline
\textit{\textbf{NS (t2-t3)}} & 0 \%       & 13 \%      & \textbf{87 \%}       & 0 \%       \\ \hline
\textit{\textbf{FH (t1-t2)}} & 0 \%       & 0 \%       & 0 \%       & \textbf{100 \%}       \\ \hline
\textit{\textbf{FH (t1-t2)}} & 0 \%       & \textbf{62 \%}       & 38 \%       & 0 \%       \\ \hline
\end{tabular}
\label{tab:details_evolution_iw}
\end{table}
%
\section{Analysis of the political opinion dataset}
\label{sec:anal_vis_int_real_data}
\begin{figure}
\centering
\includegraphics[viewport = 20 0 500 430, clip = true,  scale=0.6]{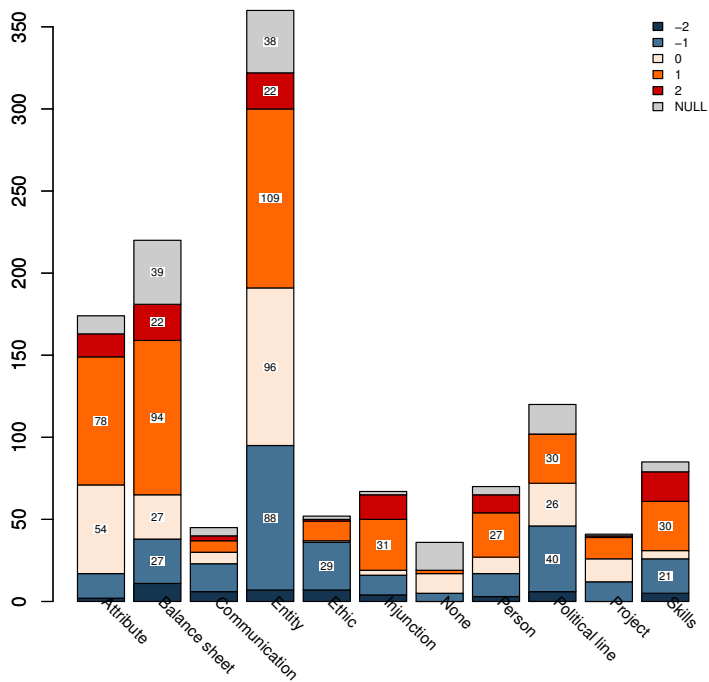} 
\caption{Illustration of the clustering results using a histogram constructed from the polarities of different aspects. The aspects are ordered from left to right as: (1) Attribute; (2) Balance sheet; (3) Communication; (4) Entity; (5) Ethic; (6) Injunction; (7) None; (8) Person; (9) Political line; (10) Project and (11) Skills. The polarities are colored and ordered from bottom to top as: -2 (dark blue), -1 (blue), 0 (light orange), 1 (orange), 2 (red) and NULL (grey).}
\label{fig:res_histogram_ex}
\end{figure}

\begin{figure*}
\centering
\includegraphics[scale=0.27]{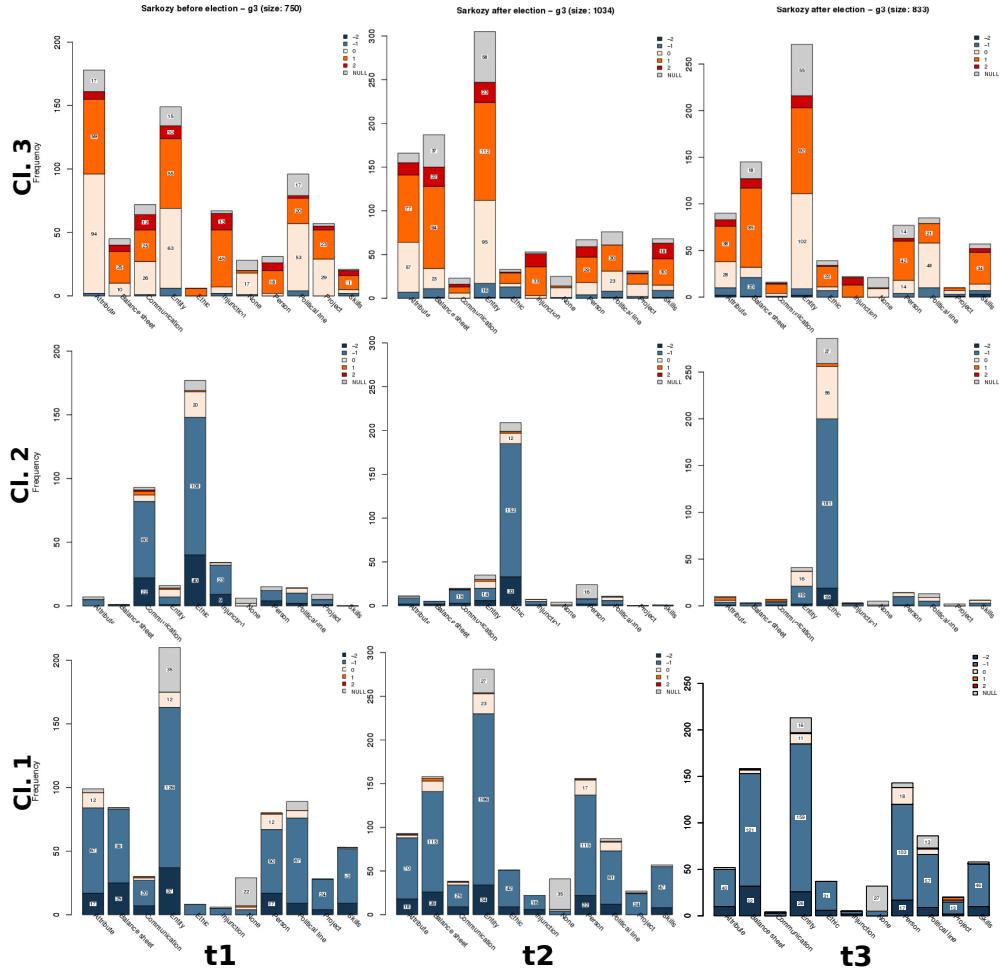} 
\caption{Illustration of the clustering results from PLMM methods for NS. Results obtained using $K=3$ for three time epochs $t1$, $t2$ and $t3$. Each cluster is represented as a histogram constructed from the polarities of different aspects. The aspects are ordered from left to right as: (1) Attribute; (2) Balance sheet; (3) Communication; (4) Entity; (5) Ethic; (6) Injunction; (7) None; (8) Person; (9) Political line; (10) Project and (11) Skills. The polarities are colored and ordered from bottom to top as: -2 (dark blue), -1 (blue), 0 (light orange), 1 (orange), 2 (red) and NULL (grey). Each column represents clusters from a particular epoch. Each row represents a particular cluster in different epochs.}
\label{fig:NS_K3_3cl}
\end{figure*}

\begin{figure*}
\centering
\includegraphics[scale=0.27]{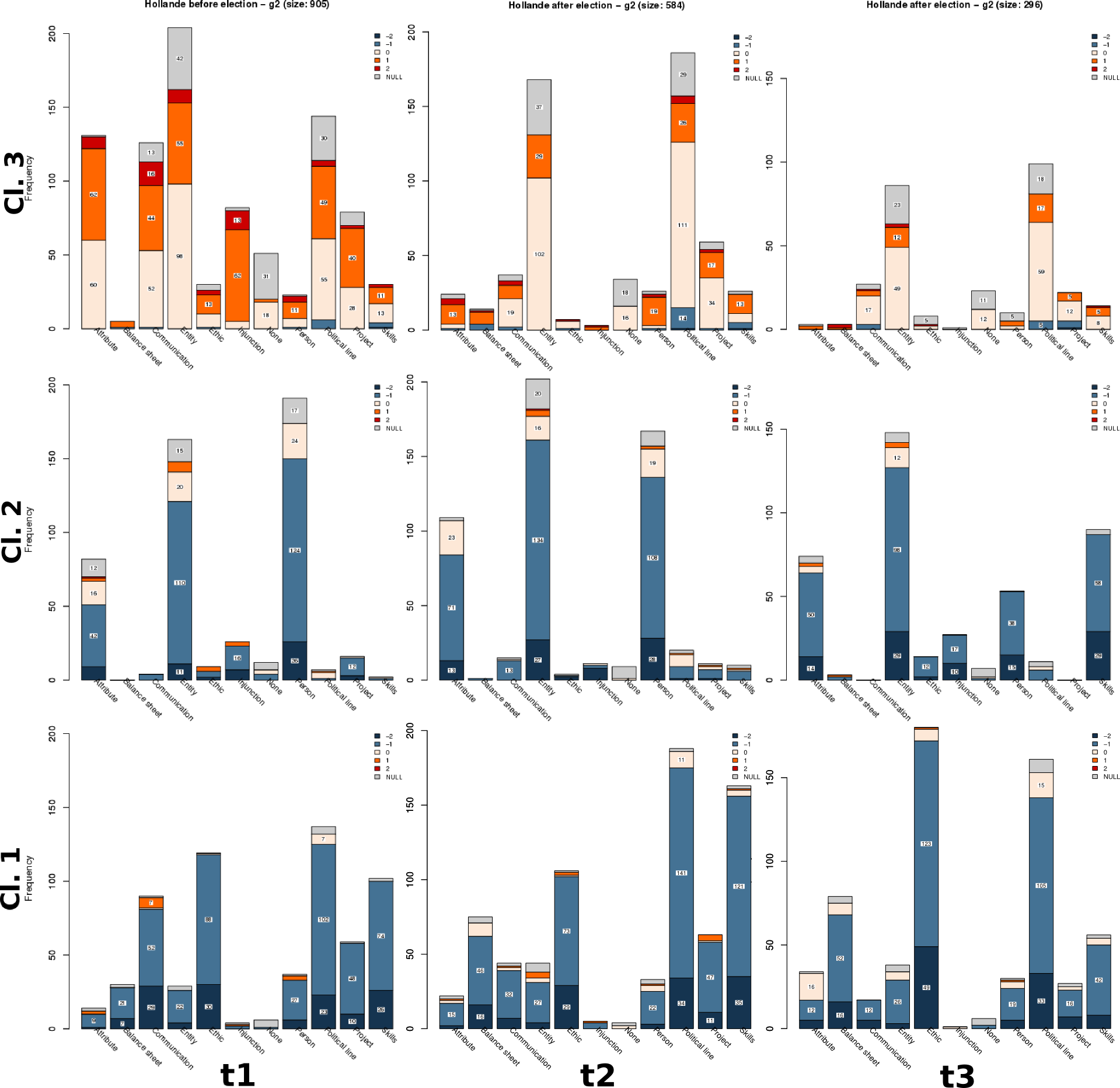} 
\caption{Illustration of the clustering results from PLMM methods for FH. Results obtained using $K=3$ for three time epochs $t1$, $t2$ and $t3$. Each cluster is represented as a histogram constructed from the polarities of different aspects. The aspects are ordered from left to right as: (1) Attribute; (2) Balance sheet; (3) Communication; (4) Entity; (5) Ethic; (6) Injunction; (7) None; (8) Person; (9) Political line; (10) Project and (11) Skills. The polarities are colored and ordered from bottom to top as: -2 (dark blue), -1 (blue), 0 (light orange), 1 (orange), 2 (red) and NULL (grey). Each column represents clusters from a particular epoch. Each row represents a particular cluster in different epochs.}
\label{fig:FH_K3_3cl}
\end{figure*}

In this section, we perform analysis on the clustering results only from the PLMM method. In order to visualize the contents, we construct a histogram representation, which helps us to discriminate among different clusters. These histograms are constructed by counting the polarities (in vertical direction) w.r.t. each attribute  (in horizontal direction). The color of the bars resembles the color of polarities. Fig. \ref{fig:res_histogram_ex} illustrates an example of a histogram which is constructed from the tweets of a cluster from time $t2$.
Following this illustration, in Fig. \ref{fig:NS_K3_3cl} and \ref{fig:FH_K3_3cl}, let us look at the examples of the clusters at different time epochs for the entities NS and FH respectively. These results are obtained by clustering data using PLMM method with $K=3$.
From both figures we observe that, at each time epoch the clusters have different histogram representations. Moreover, during different time epochs each cluster undergoes certain amount of changes in different attributes and associated polarities. This demonstrates that the proposed PLMM method is able to provide sufficient inter-cluster variations (at each time) while respecting the temporal dynamics (for each cluster during different time epochs).

\begin{figure}
\centering
\includegraphics[scale=0.27]{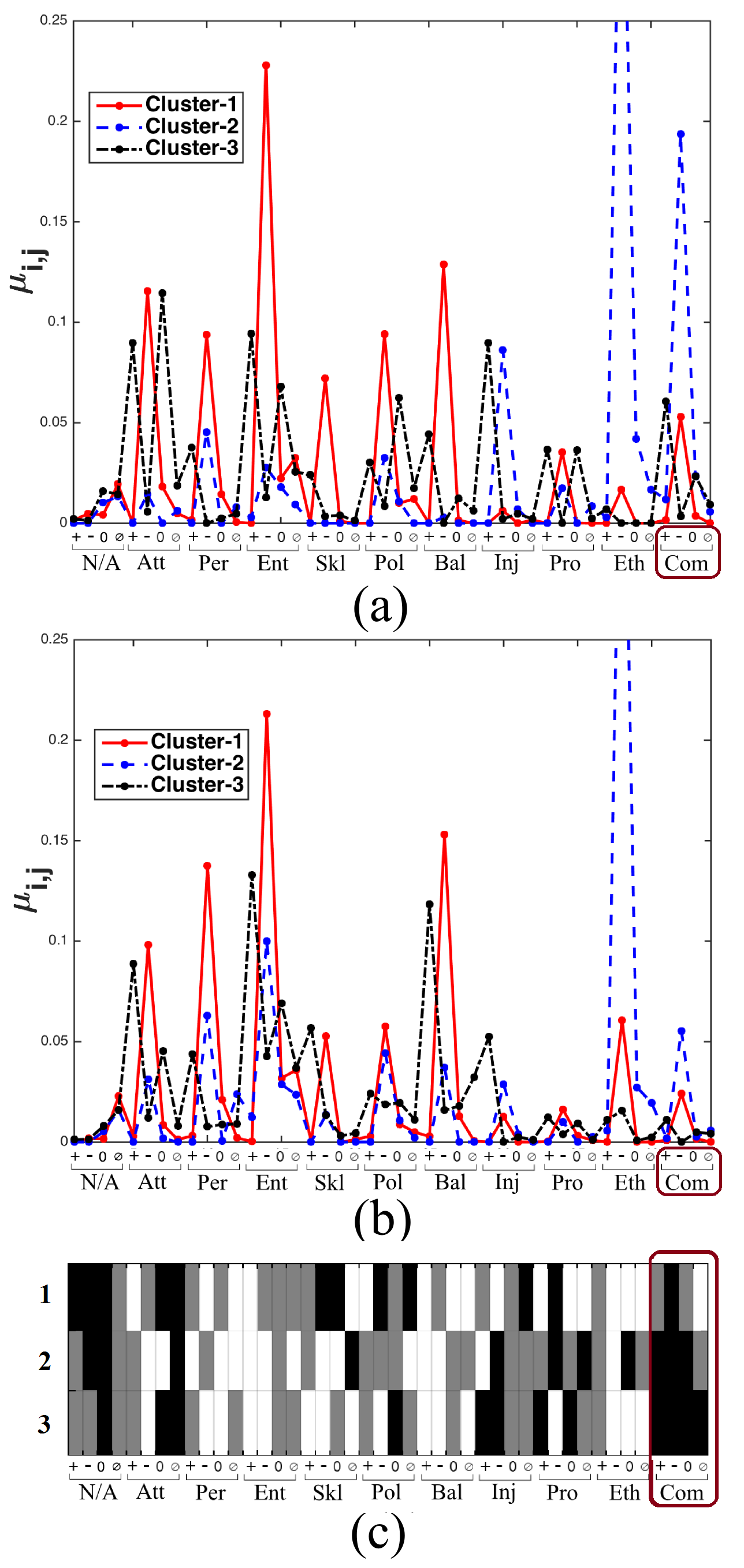} 
\caption{Example of evolution interpretation using link parameter $\delta_{k,d}$ for \textit{NS during t1 to t2 with 3 clusters}. (a) MM parameters $\mu_{k,j}^{t1}$ at time \textit{t1} (b) MM parameters $\mu_{k,j}^{t2}$ at time \textit{t2} (c) Link parameters $\delta_{k,j}$ between time \textit{t1} and \textit{t2}. In (c), for each cluster (row-wise), brighter/white color indicates the prior belief about features (aspect-polarity) increases, darker/black color indicates the prior belief about features decreases and grey color indicates the prior belief about features remains same.}
\label{fig:ex_evol}
\end{figure}

An alternative and compact representation (w.r.t. the MM model parameters) of the clusters for NS is illustrated in Fig. \ref{fig:ex_evol}(a) and \ref{fig:ex_evol}(b). 
Similar to the examples of Fig. \ref{fig:NS_K3_3cl}, this alternative representation demonstrate that, at a certain time epoch different cluster emphasizes on different aspects/polarities of an entity. Besides, the temporal changes of the clusters can be identified subsequently during different epochs by observing the increase/decrease of the probabilities.
However, from the user’s perspective, this representation may not be convenient to understand. Therefore, we use histograms for further analysis and use this compact representation for a different purpose.

Now, let us explain the semantics obtained from these clustering results. For brevity, here we denote a cluster as \textit{cl.}. From Fig. \ref{fig:NS_K3_3cl} (clusters for NS) we see that, while \emph{cl.} 1 and 3 emphasize on the negative (-) and positive (+) polarities respectively, \emph{cl.} 2 emphasizes on a particular attribute. Naively we can say that, there are three groups of peoples: (a) the first group (\emph{cl.} 1) provides negative opinions from various aspects, thus tends to hold a negative image about the entity; (b) the second group (\emph{cl.} 2) particularly emphasizes on \emph{Ethic} of the entity and mostly provide negative opinions and (c) the third group (\emph{cl.} 3) can be seen as a contrary to the first group (\emph{cl.} 1) as it tends to hold a positive image about the entity. Table \ref{tab:real_tweet_time_t1} provides three examples of the tweets for time $t1$ and for each cluster about NS. We can realize that these tweets reflect the opinions which truly correspond to the groups obtained by the clustering method.

From temporal viewpoint, we observe several changes w.r.t. different aspects. In order to analyze the changes using histograms, we observe the height of histogram bar for each aspect. This height indicates the number of tweets/opinions corresponding to the related aspect. Let us consider an example of the aspect \emph{Communication} which plays a certain role on clustering. We observe that: (a) for \emph{cl.} 1, the total number of tweets related to the aspect \textit{Communication} remains same during time $t1$ and $t2$ and reduces during $t2$ and $t3$; (b) for \emph{cl.} 2, the total number of tweets related to this aspect reduces continuously and (c) for \emph{cl.} 3, the total number of tweets related to this aspect reduces from $t1$ to $t2$ and remains same during $t2$ to $t3$. Moreover, a closer look on \emph{cl.} 3 from $t2$ to $t3$ reveals an increase of positive opinions about the \textit{communication} skill of the entity. Another example is the aspect called \emph{Attribute}, whose height reduces continuously with time for both \emph{cl.} 1 and 3. Similarly, from an analysis of the height of histogram bars in Fig. \ref{fig:FH_K3_3cl} (clusters for FH) we see that, the aspects called \emph{Entity}, \emph{Ethic}, \emph{Political line}, \emph{Skills} and \emph{Communication} play certain role to describe the image of FH. For example, the tweet - \textit{Holland would remove the word ``race'' in the Constitution (orig:  Hollande supprimerait le mot ``race'' dans la Constitution)}  from time $t1$ and \textit{cl. 3} is annotated with the aspect called \textit{political line} and polarity \textit{+1}. Another tweet - \textit{Holland and Netanyahu evoke the struggle against anti-Semitism (orig: Hollande et Netanyahou évoquent la lutte contre l'antisémitisme)} has the same annotation which is from the same cluster but from time $t3$. These  two examples reveal the importance of the aspect \textit{political line} for keeping the similar opinions into the same group at different time.
The above observations clearly indicate that, for different groups of people different aspects has certain importance at different time. Therefore, an analyst can retrieve the most prominent aspects from people’s opinion about an entity at a particular time or within a certain range of time periods.

\begin{table}[]
\centering
\caption{Real twitter data examples of the $3$ clusters at time $t1$ for entity NS. See Fig. \ref{fig:NS_K3_3cl} column 1 for the associated histograms.}
\label{tab:real_tweet_time_t1}
\begin{tabular}{|l|l|}
\hline
                & \multicolumn{1}{c|}{{ \it Cluster 1 (Generally Negative)}}                                                                                                                                                                                                                                                                                    \\ \hline
{\bf \it Ex. 1} & \begin{tabular}[c]{@{}l@{}}\textbf{Orig:} Il veut desréférendums car... y a pas de pilote dans l'avion, \\ dit-il: quel aveu! \#Sarkozy\#projet\\ \textbf{Trans:} He wants referendumbecause… there is no pilot in the plane he says: \\ what a confession! \#Sarkozy\#project\end{tabular}                                                                              \\ \hline
{\bf \it Ex. 2} & \begin{tabular}[c]{@{}l@{}}\textbf{Orig:} Je ne voterais pas \#Sarkozy ! " " Je ne voterais pas \#Sarkozy ! \\ \textbf{Trans:} I won’t vote for \#Sarkozy !" " I won’t vote for \#Sarkozy\end{tabular}                                                                                                                                                              \\ \hline
{\bf \it Ex. 3} & \begin{tabular}[c]{@{}l@{}}\textbf{Orig:}Nicolas Sarkozy, le plus mauvais président de la Vème République \\ \textbf{Trans:} Nicolas Sarkozy, the worst president of the Fifth Republic\end{tabular}                                                                                                                                                               \\ \hline
                & \multicolumn{1}{c|}{{\bf \it Cluster 2 (Negative, specially "Ethic")}}                                                                                                                                                                                                                                                                           \\ \hline
{\bf \it Ex. 1} & \begin{tabular}[c]{@{}l@{}}\textbf{Orig:} Jamais un président n'a été cerné par tant d'affaires! demain ds \\ @lematinch \#Bettencourt \#Sarkozy\\ \textbf{Trans:} Never before a president was surrounded by so many cases!” tomorrow in \\ @lematinch \#Bettencourt \#Sarkozy\end{tabular}                                                                            \\ \hline
{\bf \it Ex. 2} & \begin{tabular}[c]{@{}l@{}}\textbf{Orig:} Une liste de condamnés de l'\#UMP qui pourrait être bientôt complétée par les noms de \\ \#Sarkozy, \#Copé, \#Woerth\\ \textbf{Trans:} A list of convicted people of \#UMP soon completed by names such as \\  \#Sarkozy, \#Copé, \#Woerth \\ (the “Bettencourt case” is a famous case in which Sarkozy was involved)\end{tabular} \\ \hline
{\bf \it Ex. 3} & \begin{tabular}[c]{@{}l@{}}\textbf{Orig:} Sarkozy-Kadhafi: la preuve du financement. Et l'urgence d'une \\ enquête officielle \#affairedetat\\ \textbf{Trans:} Sarkozy-Kadhafi: the proof of funding. And the urge of an \\ official enquiry \#stateaffair \\ (Kadhafi is another case in which Sarkozy was involved in some way)\end{tabular}                              \\ \hline
                & \multicolumn{1}{c|}{{\bf \it Cluster 3 (Generally Positive)}}                                                                                                                                                                                                                                                                                    \\ \hline
{\bf \it Ex. 1} & \begin{tabular}[c]{@{}l@{}}\textbf{Orig:} N Sarkosy mots clé..challenge, défi, action, travail, réussite, formation, effort, \\ individualisation ..France Forte. Europe Forte \#NS2012 \\ \textbf{Trans:} N Sarkozy keywords..challenge, défi, action, work, success, training, effort, \\ individualization ..Strong France. Strong Europe \#NS2012\end{tabular}        \\ \hline
{\bf \it Ex. 2} & \begin{tabular}[c]{@{}l@{}}\textbf{Orig:} merci N.Sarkozy pour tout tu restera pour toujour mon Hero merci. merci \\ \textbf{Trans:} Thank you N.Sarkozy for all you will stay my hero forever thanks. thanks\end{tabular}                                                                                                                                         \\ \hline
{\bf \it Ex. 3} & \begin{tabular}[c]{@{}l@{}}\textbf{Orig:} Sarko est plus rationnel..\\ \textbf{Trans:} Sarko is more rational..\end{tabular}                                                                                                                                                                                                                                       \\ \hline
\end{tabular}
\end{table}

Besides the above interpretation of the clustering results, an analyst can obtain more information from the PLMM clustering results via the link parameters ($\delta_{k,d}$ or $\gamma_{k,d}$). After analyzing the links among MM parameters we notice that they are able to provide a compact explanation about the temporal changes during two time epochs. Fig. \ref{fig:ex_evol} illustrates an example for entity \textit{NS} from time $t1$ to $t2$ with 3 clusters, see column 1 and 2 of Fig. \ref{fig:NS_K3_3cl} for corresponding histograms. Fig. \ref{fig:ex_evol}(a) and Fig. \ref{fig:ex_evol}(b) illustrates the MM parameters (probability of aspect-polarity features) and Fig. \ref{fig:ex_evol}(c) provides a compact representation about the cluster evolutions using the values of $\delta_{k,d}$. To better understand this representation in Fig. \ref{fig:ex_evol}(c), we transform the link values as 0 (no change), -1 ($\delta_{k,d}<0.9$, belief increases) and +1 ($\delta_{k,d}>1.1$, belief decreases). In the context of the examples from the IW-POD, we can explain belief as: probability of a feature at time $t+1$ is increased from its probability at time $t$. Therefore, the belief indicates the relative significance of a particular feature w.r.t. time. An increase in the belief means that users tend to be more attracted by it. Following this, if a feature probability is nearly same at two different times then belief remains unchanged. In Fig. \ref{fig:ex_evol}, we highlight the effect of a particular aspect, called \emph{Communication} (\textit{Com}), and observe its contribution for cluster evolution. From Fig. \ref{fig:ex_evol} (a) and (b) we see that, from time $t1$ to $t2$ the probabilities are decreased mostly for \emph{cl.} 2 and 3. This means that, either the users from these clusters loose interest to discuss about \textit{Com} and focus on other aspects, or those users disappeared at time $t2$. Similar to \textit{Com}, we can observe other aspects such as \textit{Eth} (\emph{cl.} 1 and \emph{cl.} 3) and \textit{Ent} (\emph{cl.} 2 and \emph{cl.} 3) which causes cluster evolution in this example of Fig. \ref{fig:ex_evol}.

%
%

Let us analyze examples from real twitter data and observe them w.r.t. the Fig. \ref{fig:ex_evol}. If we look at \emph{cl.} 3 at time $t1$ (before election), the most likely features are often positive and it is clear that it gathers people in favor of NS. The prominent aspects are \textit{Att} (positive and neutral), \textit{Ent} (positive) and \textit{Inj} (positive), such as in the tweet - \textit{40 people @youngpop44 will be present at the great gathering in Place \#Concorde for supporting @NicolasSarkozy ! \#StrongFrance \#NS2012''}. This cluster slightly changes later at time $t2$ (just after election) towards \textit{Att} (positive), \textit{Ent} (positive) and \textit{Bal} (positive). The shift from \textit{Inj} to \textit{Bal} is clearly visible on Fig. \ref{fig:ex_evol}(c), third row: black color for \textit{Inj} means a decrease of attention whereas white color for \textit{Bal} means there are relatively more comments on the balance sheet of NS. Hence, the following message shows some nostalgia felt by many militants: \textit{Whatever the opinion of FH, NS has been a great president. FH can deconstruct all the reforms, we will never forget!}. To sum up, the $\delta$ parameter helps us to focus on what are the main changes, even though the observation could have been drawn among the other aspects. Following the same reasoning, all polarities targeting the aspect \textit{Com} are black, which proves that the performances of the politician in the media (e.g., TV, newspapers) are less important once the election is over.

Observations from numerous experiments reveal that, besides performing evolutionary clustering on the temporal data, PLMM also provide reasonable interpretation for the evolutions, thanks to the link parameters. Indeed, this clearly distinguishes PLMM from the rest of the state-of-the-art methods. Moreover, we notice that the interpretability of PLMM (using Eq. \ref{eq:init_m2}, \ref{eq:init_m3} and \ref{eq:init_m4}) can be separated out and externally plugged in with the results from any other discrete data clustering methods.
%
\section{Conclusion and Future Perspectives}
\label{sec:discussion}
Over the years, a large number of temporal data analysis methods have been proposed in several domains. 
In this paper, we only focused on the particular clustering methods which have been used for discrete data clustering and which are based on the assumption of the Multinomial distribution. 

We proposed an unsupervised method (i.e., no training from labeled data) for analyzing the temporal data. The core element of our proposal is the formulation of parametric links among the multinomial distributions. Computations of these links naturally cluster the evolutionary/temporal data. Furthermore, these links can provide interpretation for cluster evolution and also detect clusters evolution in certain cases. For experimental validation, we extensively used synthetic dataset and evaluated using the \emph{Adjusted Rand Index}. As a practical application, we applied it on a dataset of political opinions and evaluated using \emph{Perplexity} measure. Results show that the proposed method, called PLMM, is better than the state-of-the-art. Moreover, it provides an additional advantage through the link parameters in order to interpret the changes in clusters at different time. We also provide an extension of the proposed method for dealing with varying number of clusters which is not addressed by most of the recent methods.

Monitoring/tracking cluster evolution is an interesting issue which we do not explicitly and extensively manage in our proposed method, because it is not a primary objective in this paper. 
Yet, we can partially achieve this task by using certain information (parametric sub-models, see \ref{ssec:param_sub_models}) which are naturally integrated with our proposed method. That means, our proposed method can be used only as a detector of cluster evolution. At present, we consider the complete monitoring task as a future work. We believe that, an extension of several existing work can be added with our method to completely deal with this issue. For example, we can exploit\footnote{We conducted some initial experiments and found that this approach is applicable up to certain extent and should be further improved to use in our case, e.g., extend it with appropriate distance computation (e.g., using sKLD). } MEC \citep{oliveira2010mec} which is a cluster evolution monitoring method for continuous data. Besides, we can use \emph{label-based diachronic approach} \citep{lamirel2012new} by externally providing our clustering results as an input to it.

Computational complexity is a concern for the proposed method and can be considered as a limitation. From a decomposition of the computational time, we observe that most of the time is consumed by the optimization procedure (\emph{neldermead} simplex method). In future, a better optimization method can be incorporated to address this issue. Moreover, the time can be further reduced by eliminating the parametric sub-models which are experimentally found as redundant. 

Although we demonstrated the effectiveness of the proposed method only for political opinion dataset, we believe that it will be equally effective for different datasets that consist of the form of categorical data.
%
\bibliographystyle{imsart-nameyear}
\bibliography{biblio}

\end{document}